\documentclass{article}
\usepackage{arxiv}

\usepackage[utf8]{inputenc} 
\usepackage[T1]{fontenc}    
\usepackage{hyperref}       
\usepackage{url}            
\usepackage{booktabs}       
\usepackage{amsfonts}       
\usepackage{nicefrac}       
\usepackage{microtype}      
\usepackage{lipsum}
\usepackage{graphicx}
\usepackage{siunitx}
\usepackage{physics,amsmath}
\usepackage{color}
\usepackage{multirow}
\usepackage{tabularx} 
\graphicspath{ {./images/} }
\usepackage{authblk}
\usepackage{subfigure}
\usepackage {vtable}
\hypersetup{pdfborder=0 0 0}

\title{SHITARA: Sending Haptic Induced Touchable Alarm by Ring-shaped Air vortex}

\author[a]{Ryosei Kojima}
\author[b]{Akihisa Shitara}
\author[c,d]{Tatsuki Fushimi}
\author[a]{Ryogo Niwa}
\author[b]{Atushi Shinoda}
\author[a]{Ryo Iijima}
\author[a]{Kengo Tanaka}
\author[e]{Sayan Sarcar}
\author[c,d,f]{Yoichi Ochiai}

\affil[a]{Graduate School of Comprehensive Human Sciences, University of Tsukuba, Tsukuba 305-8550, Japan}
\affil[b]{Graduate School of Library, Information and Media Studies, University of Tsukuba, Tsukuba 305-8550, Japan}
\affil[c]{R\&D Center for Digital Nature, University of Tsukuba, Tsukuba 305-8550, Japan}
\affil[d]{Faculty of Library, Information and Media Science, University of Tsukuba, Tsukuba 305-8550, Japan}
\affil[e]{Birmingham City University, Birmingham City, UK}
\affil[f]{Pixie Dust Technologies, Inc., Tokyo 101-0061, Japan}

\begin{document}
\maketitle
\begin{abstract}
Social interaction begins with the other person's attention, but it is difficult for a d/Deaf or hard-of-hearing (DHH) person to notice the initial conversation cues. Wearable or visual devices have been proposed previously. However, these devices are cumbersome to wear or must stay within the DHH person's vision. In this study, we have proposed SHITARA, a novel accessibility method with air vortex rings that provides a non-contact haptic cue for a DHH person. We have developed a proof-of-concept device and determined the air vortex ring's accuracy, noticeability and comfortability when it hits a DHH's hair. Though strength, accuracy, and noticeability of air vortex rings decrease as the distance between the air vortex ring generator and the user increases, we have demonstrated that the air vortex ring is noticeable up to 2.5 meters away. Moreover, the optimum strength is found for each distance from a DHH.
\end{abstract}


\section{Introduction}
Human social interaction starts with attracting someone's attention. However, one of the challenges that d/Deaf and hard-of-hearing (DHH) people face is that it can be difficult for them to notice when someone is speaking to them. It is often difficult for people around them to recognize that they have a disability, or the severity of their disability, just by their appearance and behavior. Surrounding people may give up trying to communicate with their deaf or hard-of-hearing parents, leading to a loss of communication opportunities even within families.

Visual and haptic presentation methods are used for the accessibility approaches in notifications to DHH. In the visual presentation methods, intercom notification using light emitting diodes (LEDs) and displays is a typical example. In contrast, the haptic presentation uses smartphones and wearable devices equipped with hardware such as motors. However, it is difficult to receive notifications in visual presentation, unless the display or LEDs are within the field of view of DHH. In the haptic presentation, smartphones and wearables have the hassle and discomfort of vibrating motors, etc., that keep contact with the DHH's body and require prolonged wearing. In recent years, augmented reality (AR) glasses have been studied as a typical example, but AR glass also needs to be worn by the DHH all the time.

These problems either require DHH people to place a number of LEDs and displays within their field of vision, or they must endure the discomfort of vibrating motors and other devices, such as smartwatches and wearables, to receive haptic notifications.

\begin{figure}[t!]
    \centering
    \includegraphics[width=1.0\textwidth]{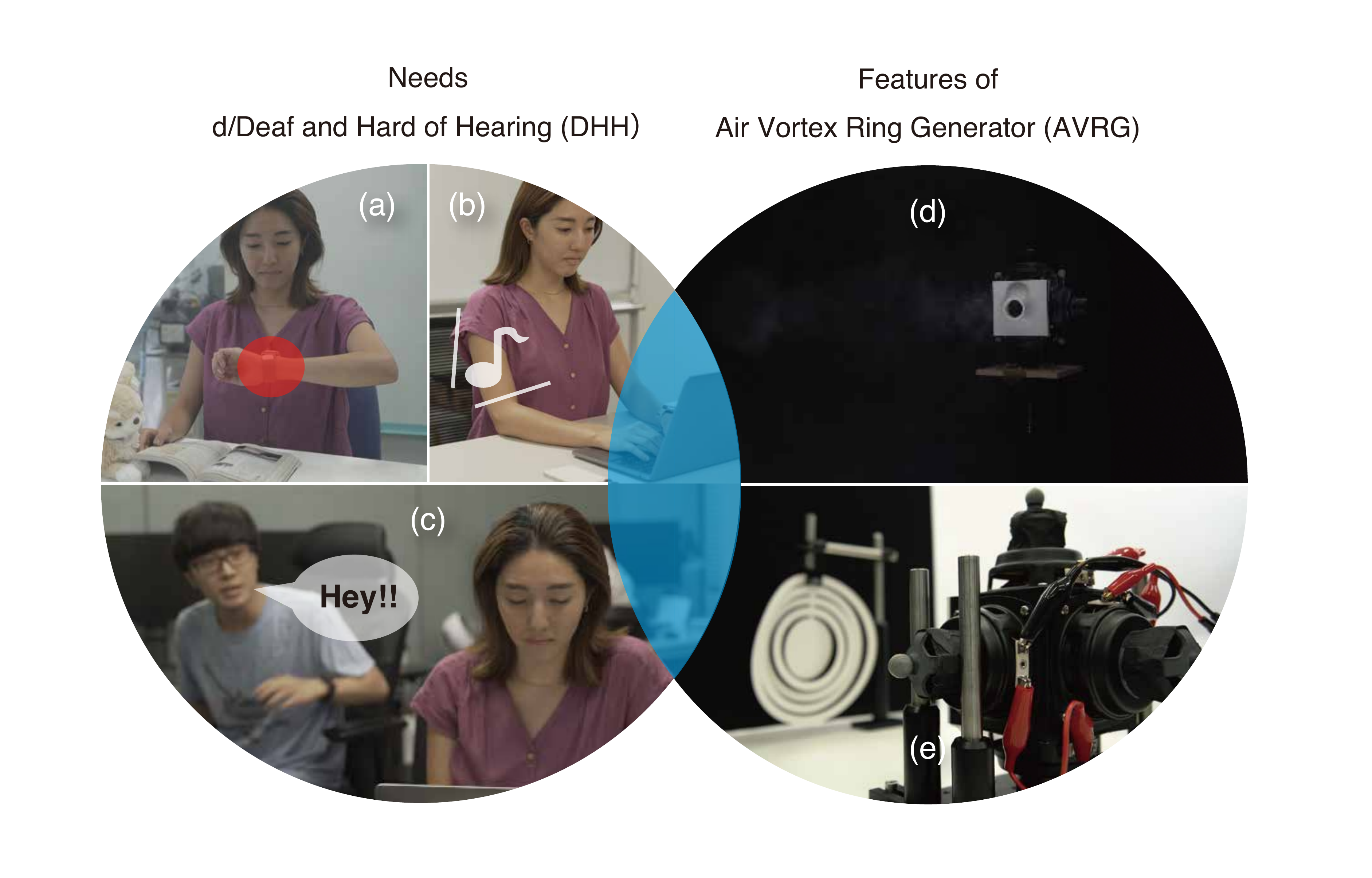}
    \caption{Feature relationship between DHH and AVRG. (a) Uncomfortable to be notified by vibration. (b) Uncomfortable to keep the devices are with the field of view. (c) Difficult to notice by voice. (d) Provides a level of sensation that is noticeable to humans over long distance (1.5m - 2.5m). (e) It would make a noise of about 70 - 84 dB to present tactile sensations over a long distance. DHH did not want to be in constant contact within the device and wanted a soft tactile sensation. AVRG is also problematic because they make noise. However, when these features are combined, it is clear that they compensate for each other.}
    \label{figure_venn_diagram}
 \end{figure}

On the other hand, there has been application research on non-contact haptic presentation, as well as application research for haptic presentation through physical contact. For example, methods using air jets\cite{Suzuki_2005}, focused ultrasound\cite{Iwamoto_2008}, and air vortex rings\cite{Sodhi_2013, Gupta_2013} have been researched and developed. Among them, AIREAL~\cite{Sodhi_2013} is a technology that allows users to feel tactile sensations in the air using air vortex rings and receive haptic feedback through interactive computer graphics and can be used in various applications such as gaming and mobile applications. AirWave~\cite{Gupta_2013} explores the use of air vortex rings to provide haptic feedback in at-a-distance interactions. Additionally, a study evaluated the effects of haptic stimuli generated by air vortex rings on the cheek on user perceptions and physiological responses, finding that different stimuli can affect these responses and task performance~\cite{Sato_2017}. Compared to other non-contact haptic presentation methods, the air vortex ring method has the following characteristics.

\begin{itemize}
     \item \textit{Advantages}
     \begin{itemize}
        \item \textit{Nothing must be prepared once the implementation is complete because it uses the atmosphere.}
        \item \textit{Provides a level of tactile sensation that is noticeable to humans over long distances (1.5 - 2.5m).}
        \item \textit{Low cost because only speakers are required.}
        \item \textit{No need to worry about wetting clothes, etc., as opposed to water-based presentation methods.}
     \end{itemize}
 
    \item \textit{Disadvantages}
     \begin{itemize}
        \item \textit{The noise level is over 70 dB.(In Fig.~\ref{figure_speaker_roundness}(b), we have achieved a quieter sound level of 46 - 53dB, but this creates a new problem: the piston speed is slower, and the momentum given is lower).}
     \end{itemize}
 \end{itemize}

Considering the characteristics of DHH people and the characteristics of the air vortex ring generator (AVRG), as shown in Fig.~\ref{figure_venn_diagram}, we propose the AVRG as a suitable initial cue of interaction with DHH. Specifically, the AVRG's feature of being able to present haptic sensations from a distance of 1.5 m to 2.5 m without keeping in touch with the user's body may be easily accepted by the DHH. AVRG has a trade-off between "sound" and "distance and momentum that can be presented," but this trade-off does not require to be considered when considering applications to DHH-only communities (e.g., residences of the DHH or Deaf space~\cite{Coates2001-Deaf, ladd2003-Deaf, padden2009-Deaf, Bartnikowska2017-Deaf}).

The use of AVRGs has not been considered in the context of accessibility as a device for talking to DHH, rather than as a graphic representation. Therefore, we first investigated the performance of the AVRG, including the audible sound generated by the AVRG, which has not been examined in detail, and how accurately the air vortex ring hits its target. As a fundamental study to establish a new accessibility approach for notifications befitting the DHH, we examined how much DHH participants were aware of the AVRG and how comfortable they felt with it and discussed the practical feasibility and improvement of the AVRG.

\begin{figure}[t!]
    \centering
    \includegraphics[width=\textwidth]{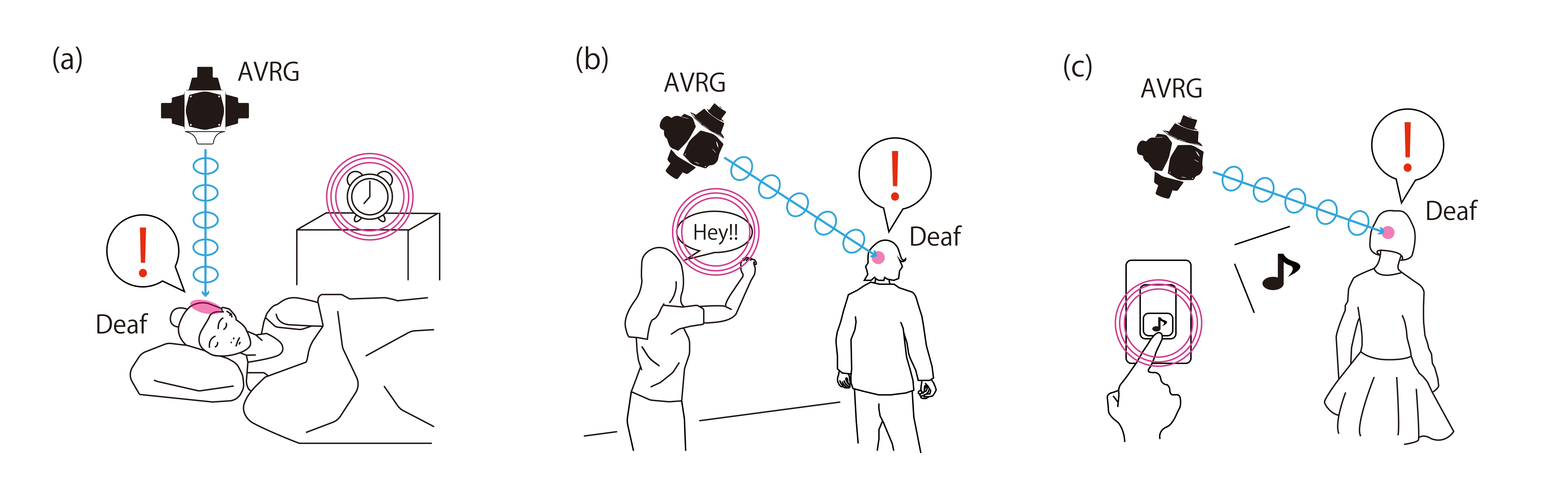}
    \caption{Notification application for possible sounds other than speech. (a) Initial conversation cue. (b) Intercom sound notification. (c) Wake-up alarm.}
    \label{fig:teaser}
\end{figure}

 As shown in Fig.~\ref{fig:teaser}, these user studies show that this method can also be used with other sound notification devices, such as alarms, intercoms, and baby monitors, to reinforce the perception of DHH people.

\section{Related Work}
\subsection{History of Support for the DHH}
The history of support for DHH shows that the first attempts were made to reinforce and convey existing auditory information. Miller Hutchison invented the electric hearing aid "Acounsticon"~\cite{World_1903} for his deaf friend in the 1900s, influenced by Bell's invention of the telephone and Edison's invention of the carbon microphone in the 1870s. The principle was to use an amplifier to take a weak signal and amplify it into a strong one. The first hearing aids were tabletop models, but later they were improved to be battery-operated, making them portable. In London, Queen Alexandria saw one of the test products and asked him to give another at Buckingham Palace, for which she gave him a gold medal. However, because of its limited frequency and dynamic range, it was not a perfect aid for the hearing impaired.

Around the 1930s, research began on converting auditory information into haptic information and delivering it to the DHH. A communication aid device for the deaf "Gault-Teletactor"~\cite{Gault_1927} was developed that divided speech into five frequency bands and mapped them to five vibrators attached to the fingers and thumb of the user's hand. The results of a study with children~\cite{Cloud_1933} showed that when the teacher pronounced "bubububububu" with the teacher's face blocked, they could not interpret the impression they got from the Gault-Teletactor. When the teacher pronounced "bububububu" while the teacher's face was visible, the children were able to connect the tactile impression with what they saw.
In the early 1950s, the "Felix" system~\cite{Tan_1997, Tan_2005} was developed by Dr. Nobert Wiener at the Research Laboratory of Electronics at MIT. It uses the cochlea model(the frequency-to-place transformation model) to convert sound frequencies into vibration locations to assist the deaf. 
In the 1990s, Tactaid $\mathrm{V}\mathrm{I}
\mathrm{I}$ ~\cite{Tan_1997, Tan_2005}, a device smaller than the "Felix" system and worn around the forearm, chest, abdomen, or neck, was developed to transmit environmental sounds through vibration and is available for commercial use. The human voice is difficult to understand, but when used in combination with lip-reading, it can be understood a little better.

When we look at the history in this way, attempts have been made to supplement auditory information that could not be received through electrical control and to convert the auditory information into vibrotactile information to be accepted by other sensory organs.

\subsection{Presentation Method for d/Deaf and Hard of Hearing}
Apple's accessibility features, ``LED flash for alerts'' \footnote{\url{https://support.apple.com/guide/iphone/led-flash-for-alerts-iph79ced06b1/15.0/ios/15.0}} and ``Sound recognition'' \footnote{\url{://support.apple.com/guide/iphone/sound-recognition-iphf2dc33312/15.0/ios/15.0}} and Google's App ``Live Transcribe \& Sound Notifications'' \footnote{\url{https://support.google.com/accessibility/android/answer/10092548?hl=en}} are probably used in the DHHs current life. Additionally, a Smart Flash Kit ``SquareGlow''\footnote{\url{https://www.squareglow.com/}} has also been developed and released by a team of DHH.
 
 Moreover, ``Ontenna\cite{Honda2022}''\footnote{\url{https://ontenna.jp/en/}} and  SoundWatch~\cite{DJ2020-SoundWatch} are known as typical examples of research. ``Ontenna'' is wearable as a presentation via vibrations and light transmission sensing with sound pressures between 60 and 90 dB. ``SoundWatch'' is an application for smartwatches that haptic and visually presents sound recognition results. Other previous studies~\cite{DJ2015-SoundAwareness,Guo2020-HoloSound} show sound's identity, location, etc., in a head-mounted display. A previous study has attempted to investigate vibrational patterns because DHH does not have desirable constant vibrational sound notifications~\cite{Goodman2020-SoundAwareness}. Another previous study~\cite{DJ2020-HoloSound} has also addressed similar watch overly persistent watch vibrations as one of the key concerns. Furthermore, in non-contact presentations other than visual and haptic for the DHH, the wasabi alarm\footnote{CHEMISTRY PRIZE: Makoto Imai, Naoki Urushihata, Hideki Tanemura, Yukinobu Tajima, Hideaki Goto, Koichiro Mizoguchi and Junichi Murakami of JAPAN, for determining the ideal density of airborne wasabi (pungent horseradish) to awaken sleeping people in case of a fire or other emergency, and for applying this knowledge to invent the wasabi alarm. \url{https://improbable.com/ig/winners/\#ig2011}}~\cite{WasabiAlarm} was researched and developed.

 Finally, other approaches outside of the DHH's daily life can also be found about haptic presentation for the DHH. In the musical scene, ``Smartphone Drum~\cite{Iijima2021-SmartphoneDrum}'' is a smartphone application that plays with haptic, presenting a drum-like vibrotactile sensation. In addition, some studies~\cite{Shitara2018-HaptStarter,Shitara2019-HaptStarter} have replaced the start signal in athletics with a haptic presentation in the sports scene. However, we cannot find any previous study on approaches to attempting non-contact haptic presentation for the DHH in accessibility.

\subsection{History of Haptic Research}
The history of tactile sensation can be traced back to the 1930s, when research was conducted to convert auditory information into tactile information, as mentioned earlier. This work was the "Gault-Teletactor"~\cite{Gault_1927}, a device that assisted DHH communication by splitting speech into five frequency bands and mapping them to five vibrators attached to the fingers and thumb of the user's hand.

In the 1940s, a master-slave manipulator~\cite{Goertz_1949, Sheridan_1989} was developed by Raymond Goertz to handle hazardous materials. The system was designed to allow a person outside the hot cell to manipulate the radioactive material contained in the hot cell, and the feel of the material as it is grasped is fed back to the user.

In the early 1950s, the "Felix" system~\cite{Tan_1997, Tan_2005} was developed. This system used the cochlea model, the frequency-to-place transformation model, to convert sound frequencies into vibration locations to assist the deaf.

In the late 1950s, a master-slave manipulator~\cite{Sheridan_1989} was developed to maintain the radioactive components of nuclear propulsion systems for aircraft. It was characterized by the fact that each finger had two degrees of freedom. In addition, a master-slave manipulator~\cite{Mosher_1964} was developed to be used for hammering nails with a hammer. However, because of the anatomical complexity of the human arm and the need to attach the actuator to the outside of the arm, it did not become very popular at the time.

In the 1960s, advances in photo electronics technology led to the development of technology that converted visual information obtained by photo sensors into tactile information such as vibration. This led to research on support for the visually impaired. For example, the Optohapt (OPtical-TO-HAPTics) system~\cite{Geldard_1966} uses nine vibration actuators attached to the body to transmit typewritten letters by vibration. TVSS (Tactile Vision Substitution System)~\cite{Bach_1969, White_1970} is a system that enables a visually impaired person to recognize objects captured by a camera (a telephone, a chair, a woman's face, etc.) using a 20 x 20 solenoid stimulus attached to the back of a chair. Gradually, as the user became accustomed to using the system, he or she was able to recognize that a telephone was in front of him or her, even if only a telephone line was captured from the camera.

In the 1980s, a portable device called Optacon (Optical-to-TActile-CONverter)~\cite{Craig_1982, Tan_2005} was developed, consisting of a hand-held camera and a 24 x 6 Braille-like tactile display. When the camera recognized the black areas of printed text, the pins would vibrate in response to the recognition.
Around this time, Internet communication between computers became possible, and the idea of having a tactile display at the end of this communication was first considered. For example, a tactile glove~\cite{Fisher_1987} for VR has been developed by NASA Ames Research Center.

In the 1990s, Project GROPE~\cite{Brooks_1990}, a tactile display using a master hand for VR, was announced. Additionally, a system~\cite{Iwata_1990} that uses a graphics computer displaying a virtual space and 9-DOF manipulators to apply reaction force to the operator's fingers and palms via the manipulators when the operator touches an object in the virtual space was also developed. The system has been demonstrated in the manipulation of solid virtual objects such as mock-ups for industrial design and 3D animated characters. PHANToM~\cite{Massie_1994}, a haptic interface, allows users to feel virtual objects by placing the device on their fingertips.
In support of research for DHH, a device called Tactaid $\mathrm{V}\mathrm{I}
\mathrm{I}$~\cite{Tan_1997, Tan_2005} has been developed, which is worn around the forearm, chest, abdomen, and neck and vibrates to transmit environmental sounds.

In the 2000s, a pin-based deformable shape display called FEELEX~\cite{Iwata_2001}, was developed. It was envisioned to be used for palpation of internal organs and 3D modeling while touching the overall shape. In the field of accessibility, research~\cite{Bach_2003} was conducted to enable the visually impaired to feel what is captured by a camera using the sense of touch on the tongue.

Later, there was research on a method of presenting tactile sensations without the device making direct contact with the body~\cite{Suzuki_2005, Iwamoto_2008, Sodhi_2013, Gupta_2013} and research~\cite{Ochiai_2016} on rendering volumetric graphics that can be touched in the air using femtosecond laser-pulsed plasma.

Thus, haptics has been used in the context of accessibility to support the deaf and visually impaired, to operate dangerous objects that cannot be handled directly by humans, and to give users the sensation of operating virtual objects such as VR and AR. However, since the birth of non-contact haptics technologies, there has been no research on the use of this technology to support DHH.
\subsection{Non-contact haptics}
First, Sensorama~\cite{Heilig_1962}, developed by Helig in 1962, is a multimodal system that includes visual presentation by video, auditory presentation by binaural sound, and tactile and olfactory presentation by breeze and odor. The breeze created by the fan can simulate riding a motorcycle without touching the viewer. Therefore, the fan is one of the examples of a non-contact tactile sensation.

Next, a portable tactile display using air jets to give localized shapes of virtual objects was developed, and the perceptual characteristics of the air jet stimulator were evaluated to see if subjects could perceive the patterns presented by this device~\cite{Amemiya_1999}. Subsequently, a tactile presentation system using air jets without rigid arms, wiring, or gloves and without physical contact was developed. However, its low resolution and the fact that it can only provide tactile sensations at a distance of approximately 30 cm were cited as problems~\cite{Suzuki_2005}.

 A focused ultrasound technique was subsequently developed~\cite{Iwamoto_2008}. This method attracted attention because it can be focused at 200 mm with 20 mm haptic resolution, and a research field of non-contact haptic presentation based on this technology was born~\cite{Carter_2013,Long_2014,Monnai_2014,Hashizume_2017}. However, the distance that can be presented by a device of approximately 200 mm square is about 400 mm, and to increase the distance, the device size must be increased, resulting in a trade-off with cost ~\cite{Hoshi_2010}.

 In addition, a method using air vortex rings was developed~\cite{Sodhi_2013,Gupta_2013}, and it is capable of presenting high pressure at a long distance of 1000 mm with a resolution of 85 mm. To the best of our knowledge, there have been two methods of AVRG developed in the field of Human-Computer Interaction, one using a speaker ~\cite{Sodhi_2013,Gupta_2013,Shtarbanov_2018,Takeda_2020} and the other using an air compressor~\cite{Sato_2017}, and both methods are known to produce audible sound.

 Additionally, a method using synthetic jets has also been developed~\cite{Shultz_2022}. This method was proposed for its advantages of relatively low cost and low voltage, but its limitation is that the reachable distance is only 25mm away.

 When comparing each of the non-contact haptics technologies, we considered AVRG's ability to be presented over a long distance as suitable for supporting talking to DHH.
 \section{Principle}
 \subsection{Air Vortex Ring}
 In this section, we will explain the principles of air vortex rings and what parameters are essential for a non-contact haptic cue for DHH.
 
\subsubsection{Principle of air vortex ring generation}
\begin{figure}[t!]
    \centering
    \includegraphics[keepaspectratio, width=0.8\textwidth]{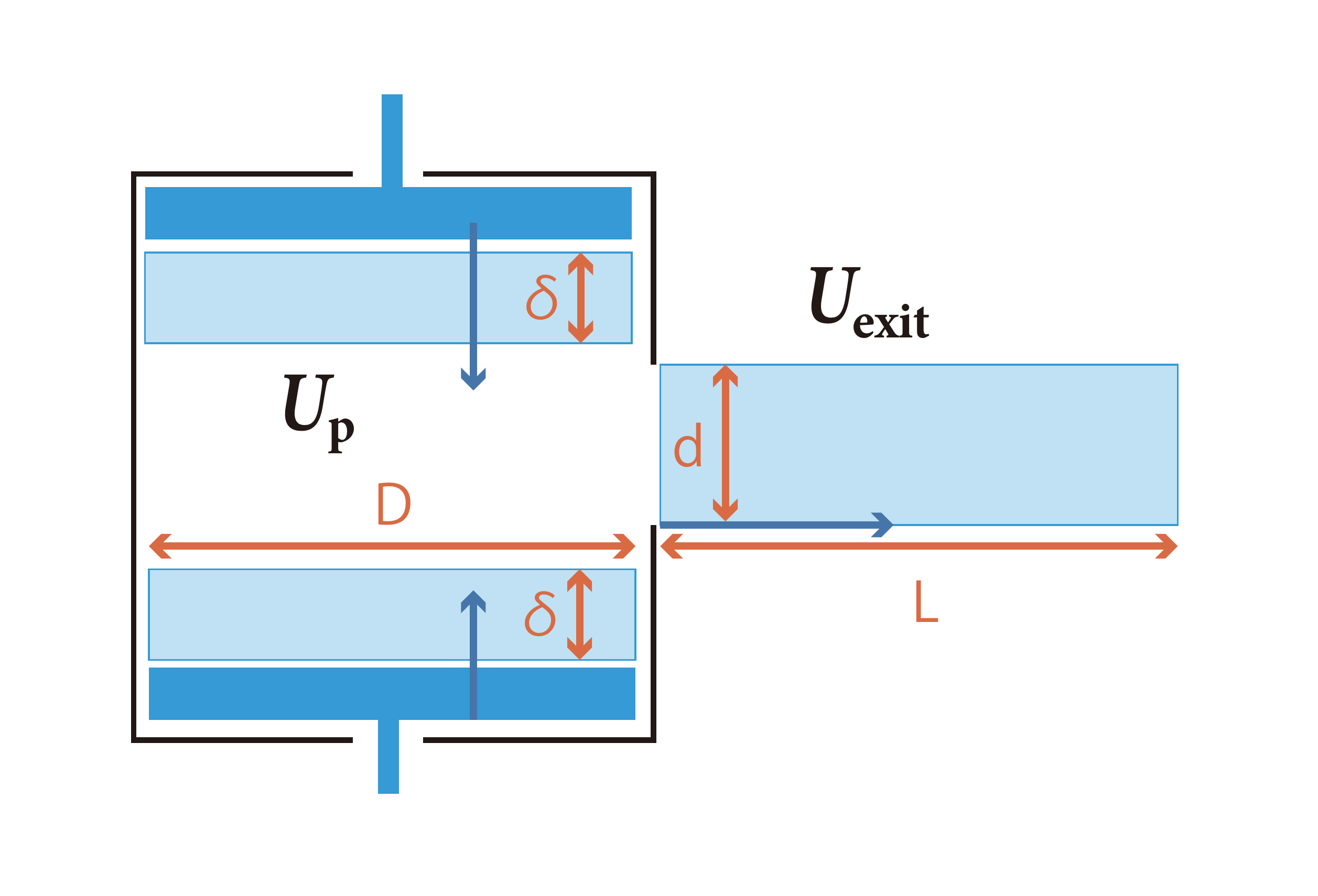}
    \caption{A cross-sectional view of an AVRG. $D$ is piston diameter, $\delta$ is piston displacement, $d$ is aperture diameter, $L$ is the length of a slug of air pushed out of the nozzle, $U_{p}$ is the piston velocity and $U_{exit}$ is the flow velocity at the aperture.}
    \label{figure_air_vortex_generator_principle}
 \end{figure}

Air vortex rings are torus-shaped air vortices that can travel long distances through the air and exert momentum on the surface they collide with. Fig.~\ref{figure_air_vortex_generator_principle} shows a cross-sectional view of an AVRG. 

 Assuming that air is incompressible, the volume of air pushed out by a piston $V_{p}$ is equal to the volume of air ejected through the opening $V_{exit}$
 
  \begin{equation}
    V_{p} = n\frac{\pi D^2}{4}\delta = V_{exit} = \pi\frac{\pi d^2}{4}L, 
  \end{equation}
  where $n$ is the number of pistons, $D$ is the piston diameter, $\delta$ is the piston displacement, $d$ is the aperture diameter, and $L$ is the length of a slug of air pushed out of the nozzle. Dividing both sides by $d$, the ratio of length to diameter $L/d$ can be summarized as
  
  \begin{equation}
    \frac{L}{d} = \frac{n\delta D^2}{d^3}.
  \end{equation}
 
 The formation number $f$ is defined as the limiting value of $L/d$ when only a vortex ring is formed without a turbulent wake and lies in the range of 3.6 - 4.5 for a broad range of flow conditions~\cite{Gharib_1998}. Therefore, $d$ is expressed as
 
  \begin{equation}
    d = \sqrt[3]{\frac{n \delta D^2}{f}}.
    \label{aperture_diameter}
  \end{equation}
 In our study, we designed the aperture diameter $d$ with a formation number of 3.85 - 4.03.
 
 Assuming that the flow rate of air pushed out by the piston $Q_{p}$ is equal to the flow rate of air ejected through the opening $Q_{exit}$~\cite{Krueger_2005,Krueger_2008}:
 
  \begin{equation}
    Q_p = n\frac{\pi D^2}{4}U_p = Q_{exit} = \frac{\pi d^2}{4} U_{exit},
  \end{equation}
 where $U_{p}$ is the piston velocity and $U_{exit}$ is the flow velocity at the aperture. To summarize,
 
  \begin{equation}
    U_{exit} = \frac{nD^2}{d^2} U_p.
  \end{equation}
 The momentum $I$ of the air vortex ring is expressed as
 
 \begin{equation}
    \begin{split}
    I &= \frac{1}{4} \pi d^2 \rho L U_{exit} \\
      &= \frac{1}{4} \pi d^2 \rho (f d) \frac{n D^2}{d^2} U_p \\
      &= \frac{1}{4} \pi \rho n f d D^2 U_p,
    \end{split}
    \label{Monmentum of AVRG}
 \end{equation}
 where $\rho$ is the density of air. In this study, we focused on the fact that the piston velocity is proportional to the momentum of the air vortex ring without changing the structure of the AVRG.

\subsubsection{Principle of silencing audible sound during air vortex ring generation}
 In the field of human-computer interaction, there have been several air vortex ring generation methods using speakers~\cite{Sodhi_2013,Gupta_2013} and air compressors~\cite{Sato_2017,Yanagida_2004}. However, the problem with both methods is that they produce audible noise during ejection. Because the system of reduced noise still desirable, we attempted to reduce the noise by rounding the square waveform, which is the input waveform, to the speaker. The square waveform was filtered with a rational transfer function, which provides the input-output description of the filter operation on a vector in the Z-transform domain, as follows:
 
 \begin{equation}
    Y(z) = \frac{b}{1 - (b-1)z^{-1}} X(z),
    \label{filter operation}
 \end{equation}
 where $X(z)$ is the input signal, $Y(z)$ is the output signal, and $b$ is the roundness coefficients of the speaker waveform. Fig.~\ref{figure_speaker_roundness} (a) shows the filtered waveforms with $b=0.001, 0.002, 0.003, 0.004$, and $1$, respectively.

\begin{figure}[t!]
    \centering
    \includegraphics[keepaspectratio, width=0.7\textwidth]{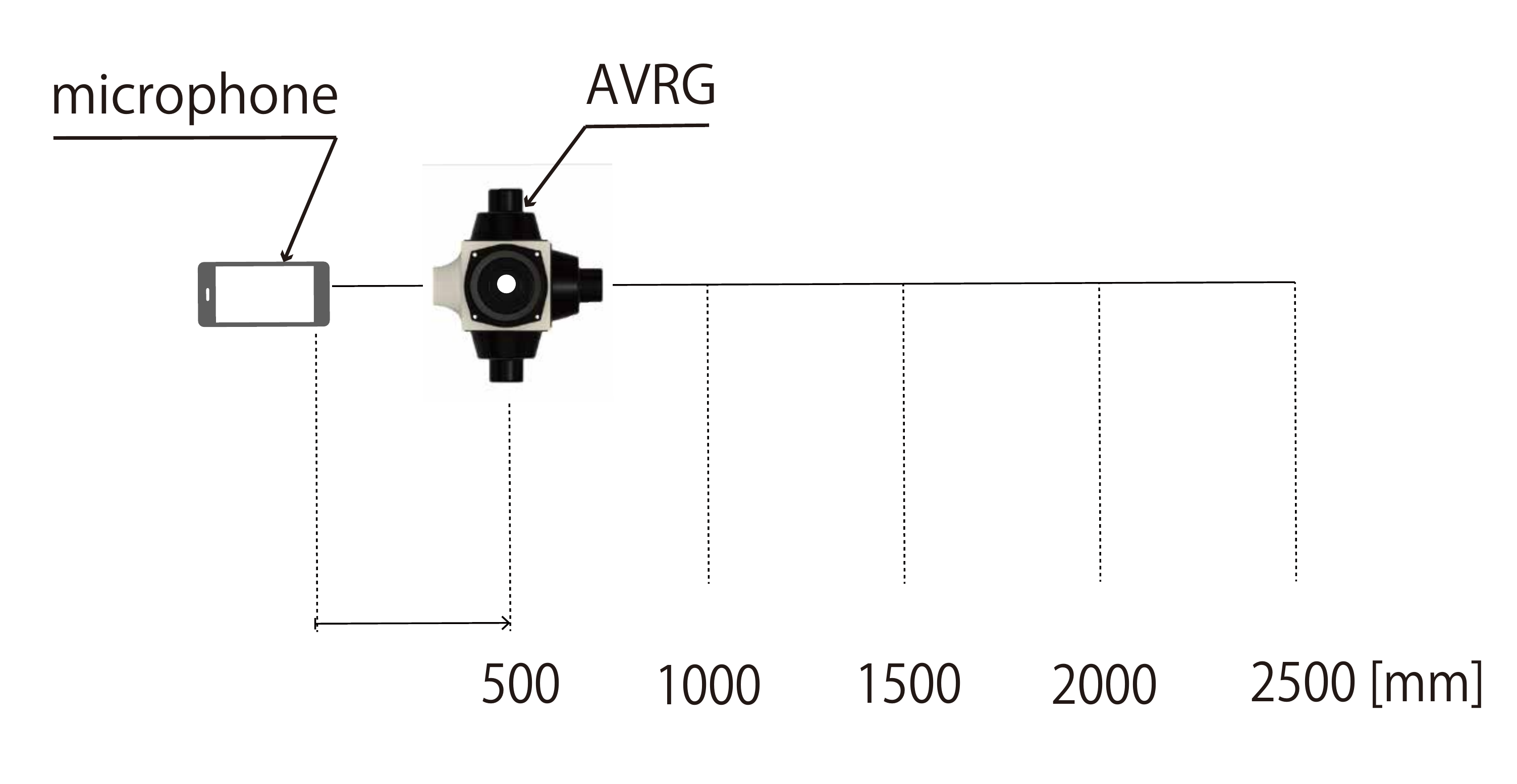}
    \caption{Noise Measurement Methods. The distances between the center of the AVRG and the microphone were 500 mm, 1000 mm, 1500 mm, 2000 mm, and 2500 mm apart.}
    \label{figure_measurement_noise}
\end{figure}
 
 \subsubsection{Measurement experiment of the sound generated by the AVRG}
The maximum sound pressure levels were measured when these waveforms were driven; the center of the AVRG and the microphone height were aligned at 140 mm each, and the distances were 500 mm, 1000 mm, 1500 mm, 2000 mm, and 2500 mm apart, as shown in Fig.~\ref{figure_measurement_noise}. The iPhone application NIOSH (National Institute for Occupational Safety and Health) Sound Level Meter (EA LAB, Slovenia), which was found to be the most effective of the nine apps in the previous study~\cite{Crossley_2021}, was used for the measurements, and the device was an iPhone 12 Pro. The sound generated by AVRG is classified as an isolated burst of sound energy, and the maximum value of each burst is generally constant. Therefore, the maximum instantaneous noise level within 10 measurements was used as the data. The ambient temperature was 24.2 °C (75.56 °F).

\begin{figure}[t!]
    \centering
    \subfigure{\includegraphics[keepaspectratio, width=0.48\textwidth]{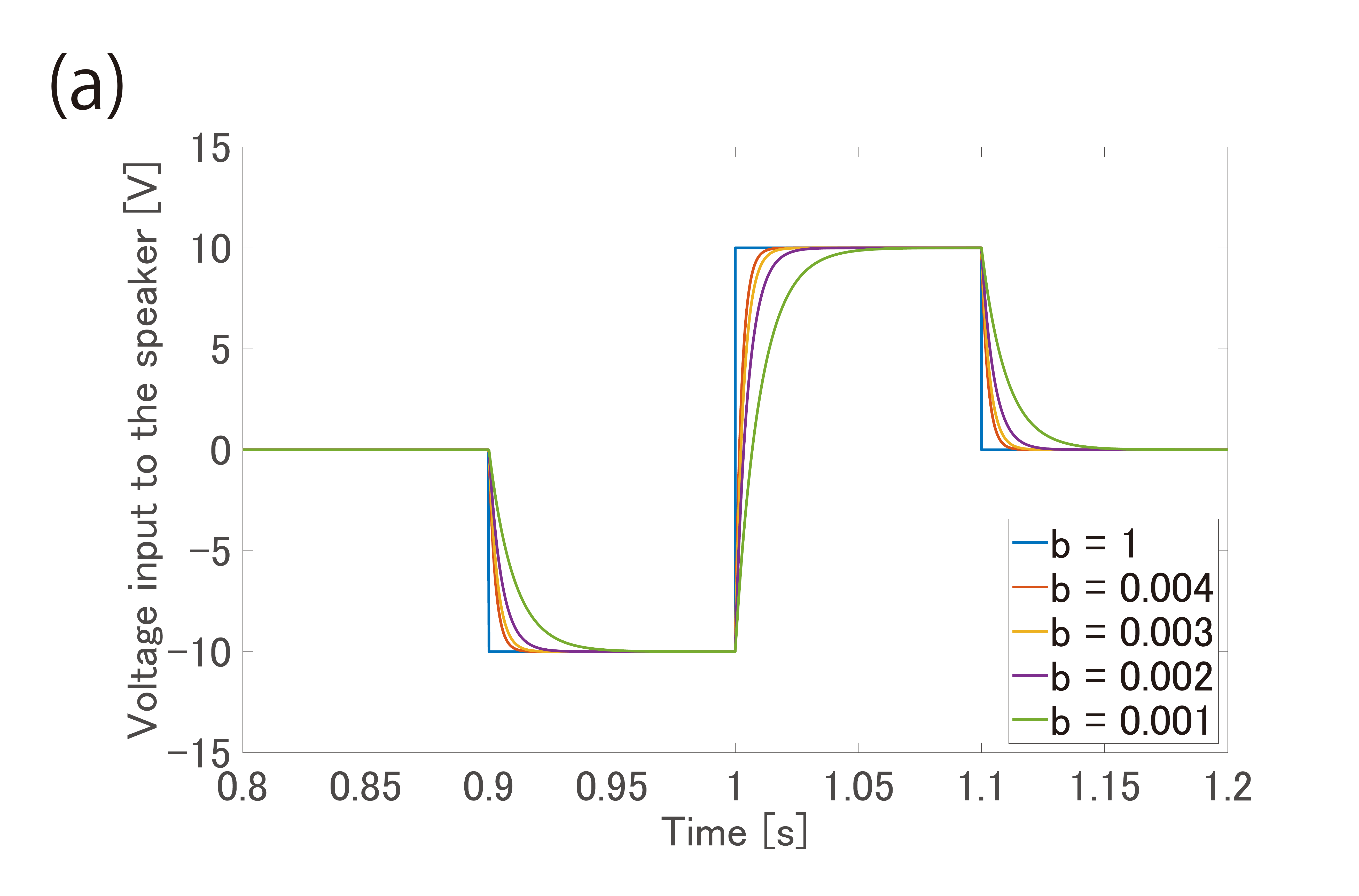}}
    \subfigure{\includegraphics[keepaspectratio, width=0.48\textwidth]{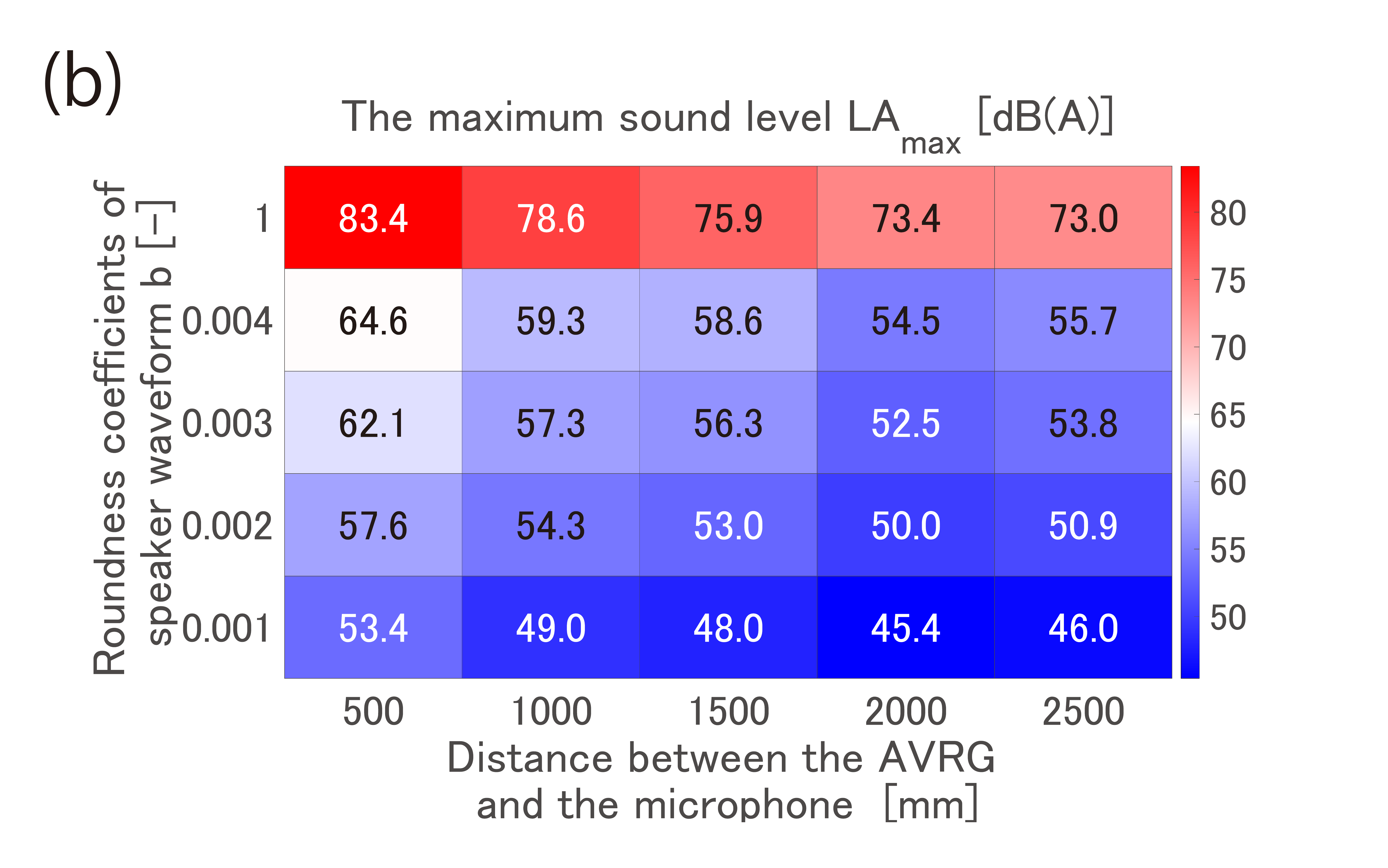}} 
    \subfigure{\includegraphics[keepaspectratio, width=0.48\textwidth]{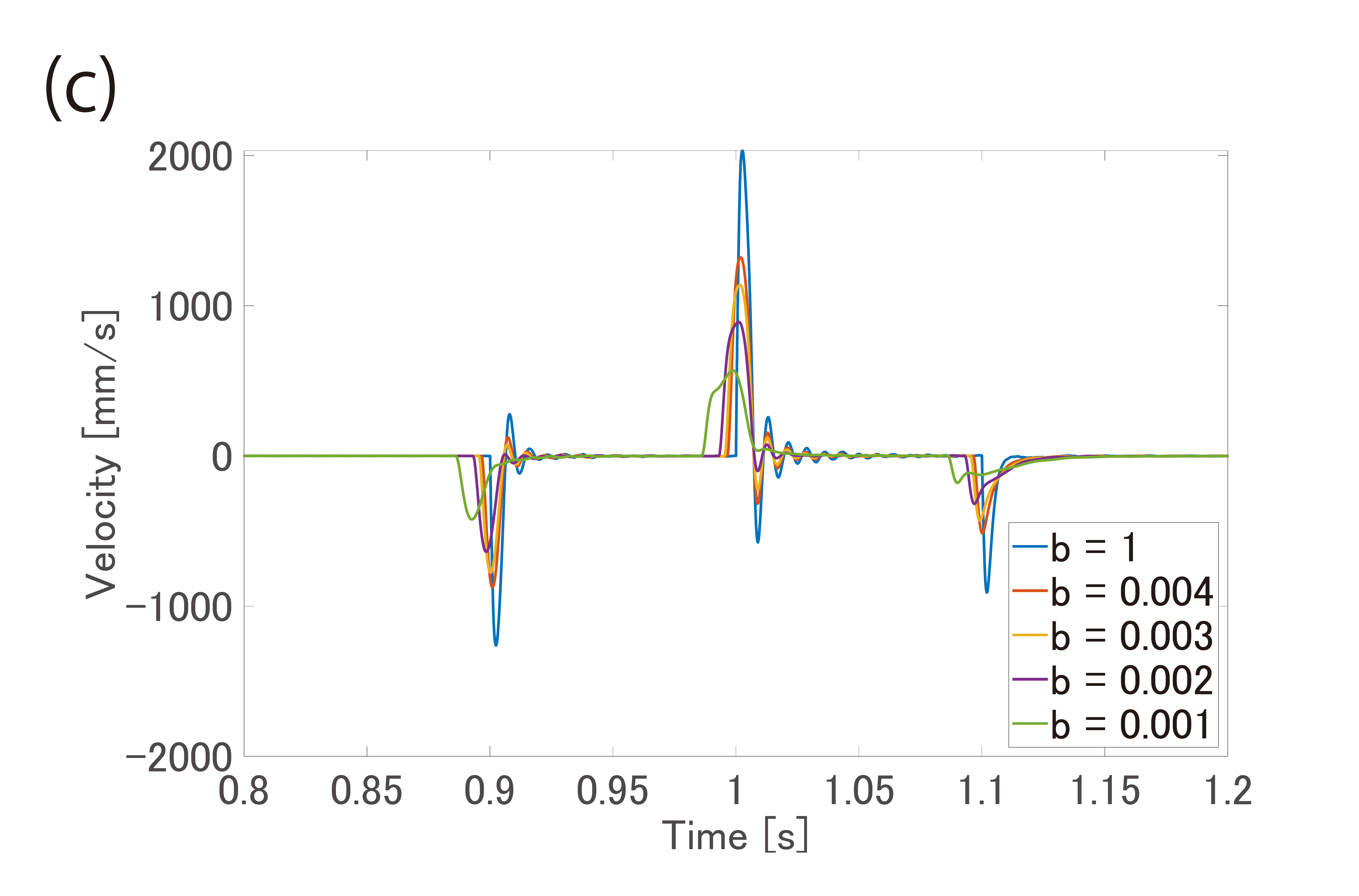}}
    \subfigure{\includegraphics[keepaspectratio, width=0.48\textwidth]{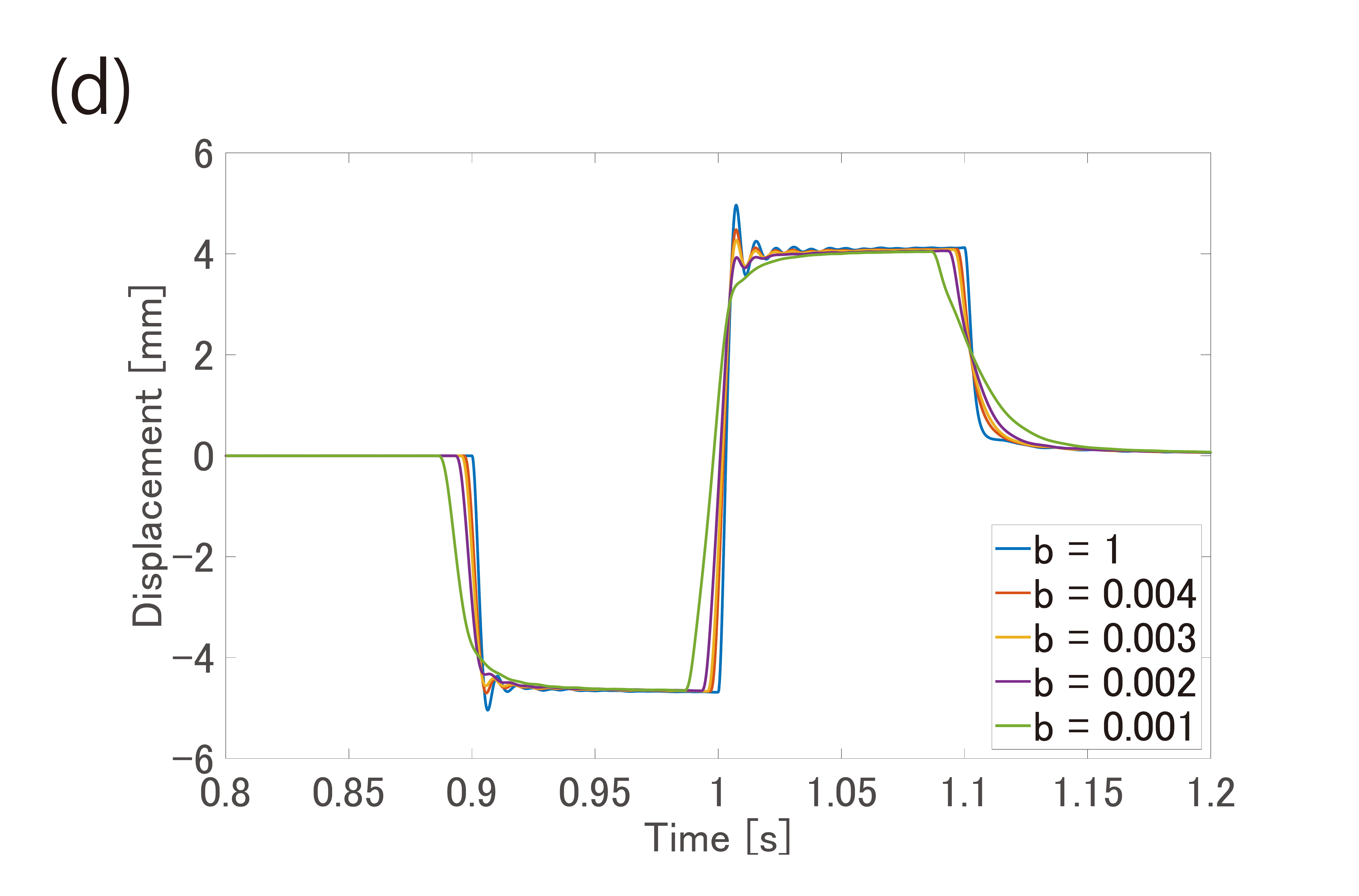}} 
    \caption{Principle of silencing audible sound during air vortex ring generation and measurements of the movement of the speaker membrane with a scanning vibrometer (PSV-500, Polytec) when the waveform in Fig.~\ref{figure_speaker_roundness} (a) is input. (a) Voltage waveforms input to the speaker after filtering with roundness coefficients, as shown in Equation \ref{filter operation}. (b) The maximum sound level $LA_{max}$ when the voltage waveform is input. The more the input waveform of the speaker is rounded, the quieter the sound becomes for all distance conditions measured.(c) Speaker velocity. The maximum velocities are 568.83 mm/s, 892.97 mm/s, 1139.30 mm/s, 1321.72 mm/s, and 2032.20 mm/s, respectively.(d) Speaker displacement. The maximum value is 8.6mm(-4.6 - 4.0mm).}
    \label{figure_speaker_roundness}
\end{figure}
 
  Fig.~\ref{figure_speaker_roundness} (b) shows the more the input waveform of the speaker is rounded, the quieter the sound becomes for all distance conditions measured. Additionally, the maximum sound pressure generated was lower when the roundness coefficients were reduced than when the distance from the AVRG was increased. However, when the piston velocity (velocity of speaker membrane) was measured using a scanning vibrometer (PSV-500, Polytec) as shown in Fig.~\ref{figure_speaker_roundness} (c), we also observed reduction in the piston velocity as a consequence, which limits the momentum of the air vortex ring. Comparing the case with no rounding (b = 1) and the case with a slight rounding (b = 0.004), the maximum sound pressure was reduced by approximately 20 dB while the loudspeaker speed was reduced by approximately half. This is equivalent to a factor of 10 in terms of sound pressure, which shows that the system contributes to a reduction of sound pressure. This experiment allowed us to verify how much the sound generated by the speaker changes relying on the roundness coefficients of the square wave input to the speaker.
\section{Implementation}
\subsection{AVRG System}
\begin{figure}[t!]
    \centering
    \includegraphics[keepaspectratio, width=0.8\textwidth]{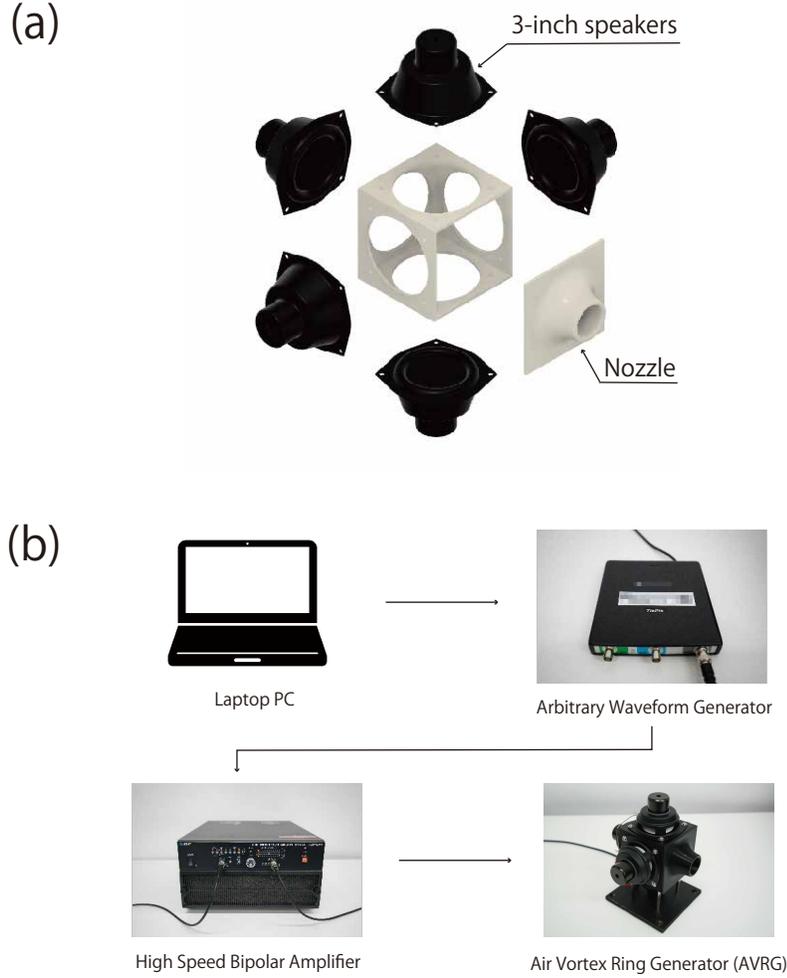}
    \caption{(a)AVRG Configuration. The five 3-inch speakers (NS3-193-8A, AURA SOUND) are mounted around the enclosure of the AVRG and serve as a piston. (b)System Overview. The system consists of a PC, a USB oscilloscope capable of generating arbitrary waveforms (Handyscope HS5, TiePie engineering), a high speed bipolar amplifier (HSA42014, NF Corporation), and an AVRG.}
    \label{figure_implementation_overview}
 \end{figure}

\begin{figure}[t!]
    \centering
    \includegraphics[keepaspectratio, width=0.8\textwidth]{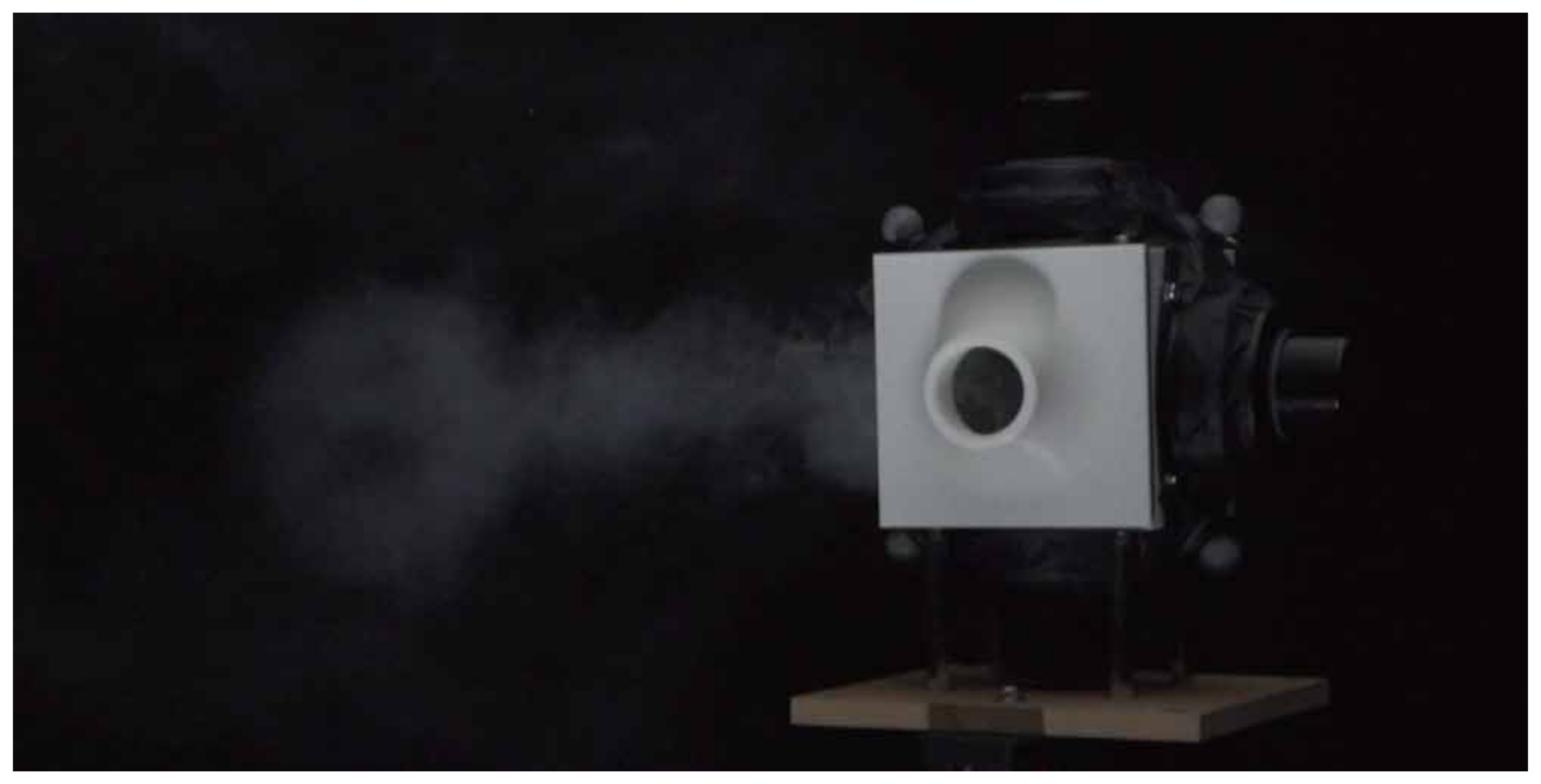}
    \caption{Smoke visualization of an air vortex ring ejected from this AVRG.}
    \label{figure_AVRG_smoke}
  \end{figure}

 Fig.~\ref{figure_implementation_overview} shows an overview of the system: a PC, a USB oscilloscope capable of generating arbitrary waveforms (Handyscope HS5, TiePie engineering), a high speed bipolar amplifier (HSA42014, NF Corporation), and an AVRG. The AVRG comprises a 3D printed enclosure (94.5mm x 94.5mm x 94.5mm), five 3-inch speakers (NS3-193-8A, AURA SOUND) with a rated impedance of 8 $\Omega$, and a 3D printed nozzle with a 30mm diameter aperture. The five speakers are mounted around the enclosure of the AVRG and serve as a piston. As shown in  Fig.~\ref{figure_speaker_roundness} (d), when a filtered voltage waveform of 10 V is driven to the speaker, the displacement is approximately 8.2 mm. Based on  Equation~\ref{aperture_diameter}, the number of speakers $n$=5, the speaker diameter $D$=51.5[mm] and the information number $f$=4.03 were determined and the aperture diameter was set to 30mm.
 The overall size of the AVRG is 204.9 mm x 204.9 mm x 185.6 mm, with a total weight of 1.308 kg.
  Fig.~\ref{figure_AVRG_smoke} shows a smoke visualization of an air vortex ring ejected from this AVRG.

  \begin{figure}[t!]
    \centering
    \includegraphics[keepaspectratio, width=1.0\textwidth]{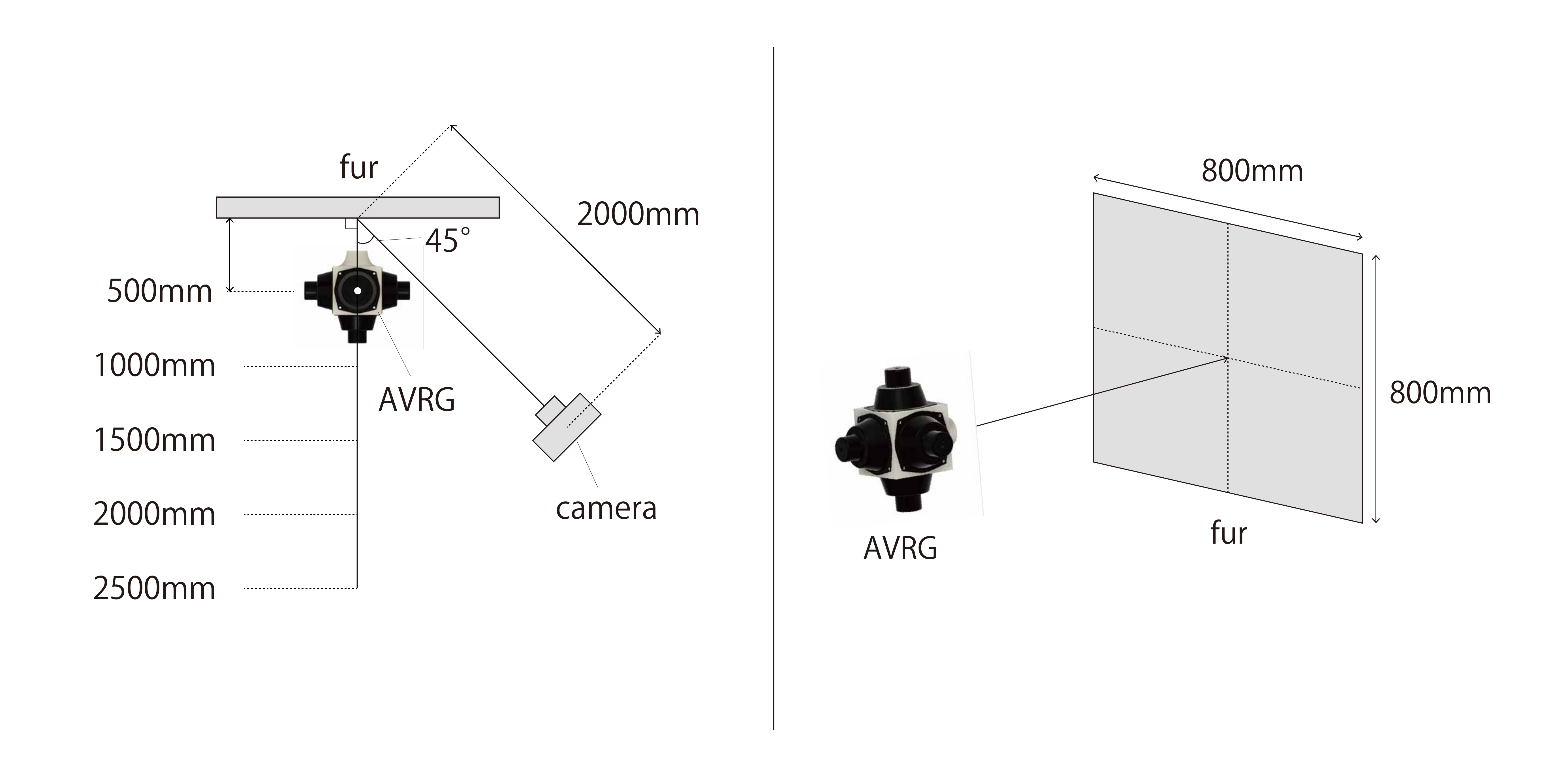}
    \caption{Overview of accuracy experiment. Experiments were conducted under five conditions with AVRG and fur distances of 500, 1000, 1500, 2000, and 2500mm. A camera was placed horizontally at an angle of 45° from the center of the fur at 2000 mm to photograph the swinging of the fur.}
    \label{figure_overview_of_fur_experiment}
  \end{figure}

\subsection{Accuracy Experiment}
 Displays have been proposed that utilize the phenomenon that fur shading changes when fur fibers are made to stand up or flatten using a roller device, pen device, or focused ultrasonic device~\cite{Sugiura_2014}. The discrepancy between the aimed location of the air vortex ring and the actual location of the air vortex can be visualized using fur. When an air vortex ring strikes a fur material, the stroked location instantly shakes. We measured the shift of the air vortex ring from the center of the target fur relying on the location where the fur was shaken and validated how accurately the air vortex ring hits its target.

 \begin{figure}[h!]
    \centering
    \includegraphics[keepaspectratio, width=1.0\textwidth]{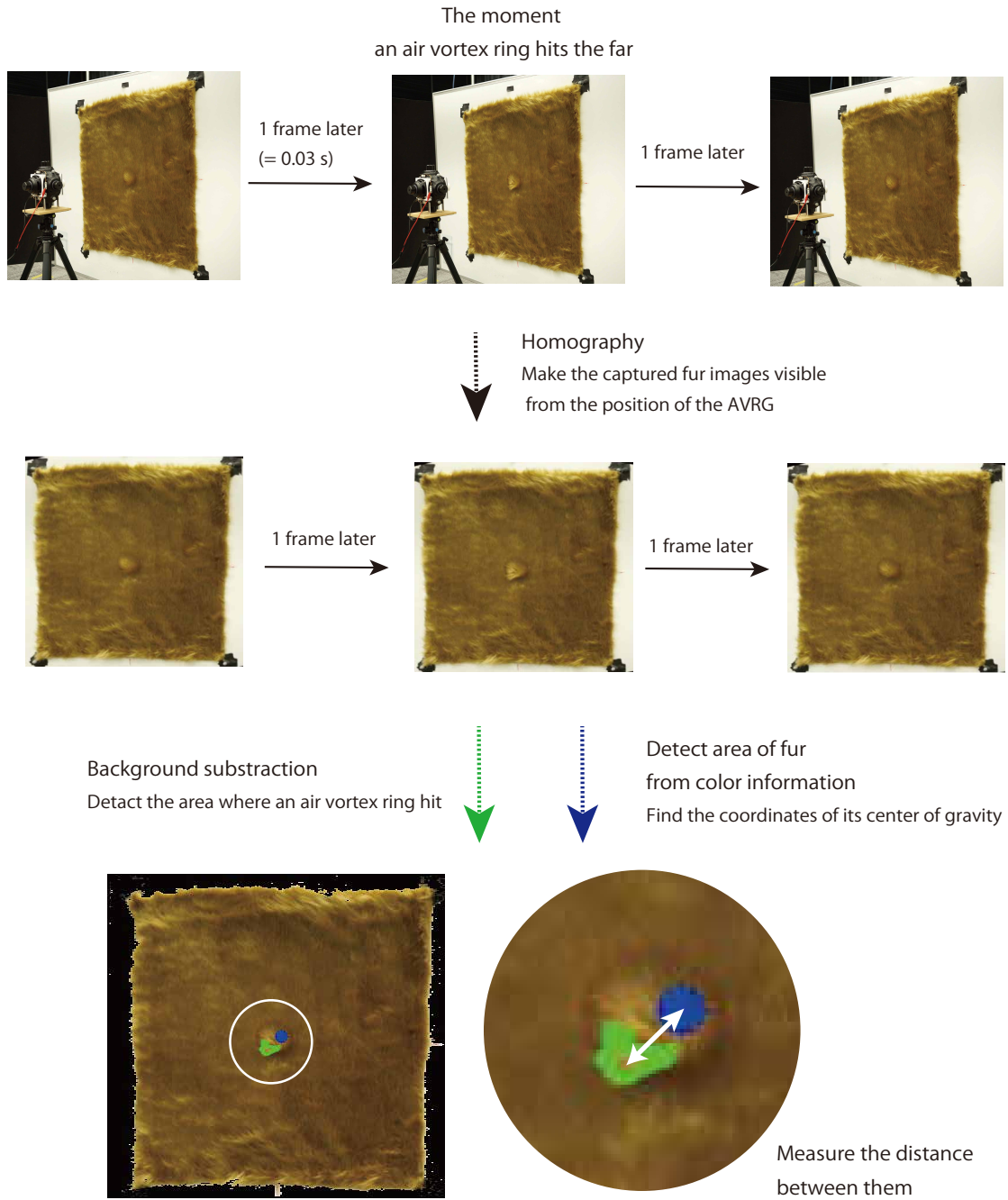}
    \caption{Fur video analysis process. A homography was performed to make the captured fur images visible from the position of the AVRG. The coordinates at which the air vortex ring shook the fur were identified by the background subtraction method. The entire area of the fur was identified from the color information of the fur. The center coordinates were identified, and the distance between them was measured.}
    \label{figure_fur_analysis_process}
  \end{figure}
 
\subsubsection{Experimental Environment and Condition}
  Fig.~\ref{figure_overview_of_fur_experiment} shows an overview of the experiment. The fur area was 800 mm × 800 mm, and the fur length was about 40 mm. An AVRG was placed perpendicular to the plane of the fur at the same height as the center of the fur with motion capture cameras (seven Prime17W, two PrimeX22, and one Prime41, OptiTrack). A camera was placed horizontally at an angle of 45° from the center of the fur at 2000 mm to photograph the swinging of the fur. The AVRG did not overlap with the fur plane because of the camera placed in this position. 
  
 A homography was performed to make the captured fur images visible from the position of the AVRG, as shown in Fig.~\ref{figure_fur_analysis_process}. The center of the fur, which is the target of the air vortex ring, was obtained as the coordinates of the center of gravity of the entire fur region. The area hit by the air vortex ring was detected using background subtraction, and its center of gravity was determined as the fur hit position.

 The experimental conditions were set at five different distances between the AVRG and the fur, rearranging from 500-2500 mm, with five different roundness coefficients of speaker waveform: 0.001, 0.002, 0.003, 0.004, and 1, for a total of 25 patterns. An air vortex ring was applied 20 times to the fur, and the measurement data were randomly extracted 10 times, relying on the number of times a detection was possible, and the mean value was taken.

\subsubsection{Result and Discussion}
\begin{figure}[h!]
    \centering
    \includegraphics[keepaspectratio, width=1.0\textwidth]{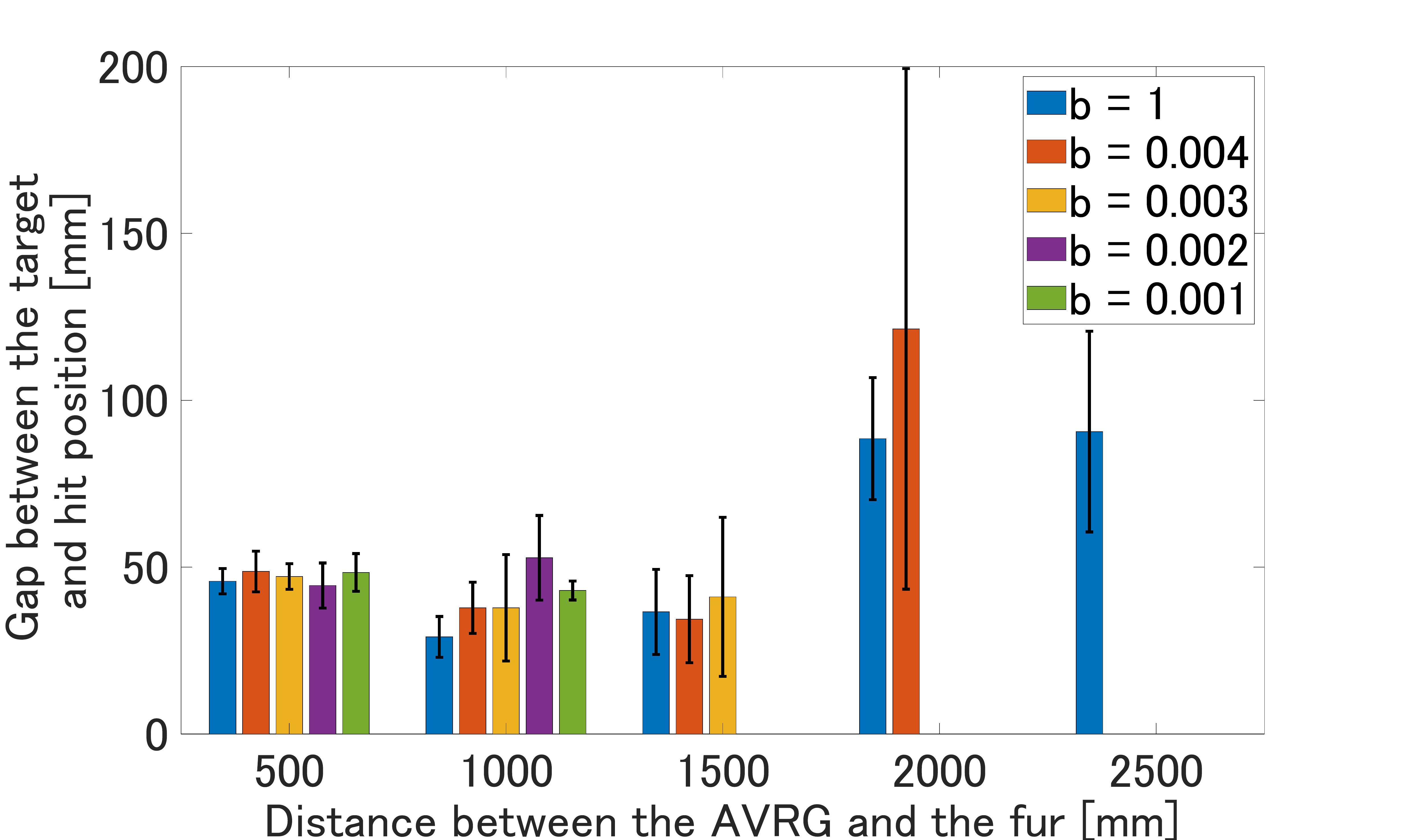}
    \caption{Result of accuracy experiment. The bar represents the mean of the gap between the target and the hit position, and a vertical error bar represents its standard deviation.
    The data without bars was considered N/A because the fur motion could not be detected by the image analysis because of weak swaying.}
    \label{figure_result_of_fur_experiment}
  \end{figure}

  Fig.~\ref{figure_result_of_fur_experiment} shows the results. The data without bars were considered N/A because the fur motion could not be detected by the image analysis because of weak swaying. Because only a portion of the area hit by the air vortex ring could be detected, the center coordinate within the range where the air vortex ring motion was detected is the point where the air vortex ring hit. The radius of the air vortex ring (50 mm) is considered the error range.
 Therefore, the target coordinates were primarily included in the area where the vortex ring hit the target when the distance was between 500mm and 1500mm. When the distance was more significant than 2000 mm, the gap between the air vortex ring and the target became larger. However, the mean head breadth, head thickness, and mention-top of head for Americans were 145 mm, 194 mm, and 241 mm, respectively ~\cite{Lee_2006}, suggesting that it was possible to hit the DHH person's head in most patterns. 

\section{User Study}
To quantitatively evaluate at what level DHH noticed and qualitatively how the noticeability and comfortability by DHHs changes by various parameters, we performed two user studies.  In User study 1, we varied the distance (between head and AVRG) and roundness coefficients of the speaker waveform. User study 2 varied the direction of the AVRG.  At the end of the experiment, we asked the participants to fill out a questionnaire to obtain qualitative feedback. The mean temperature of the experimental environment was 26.1°C (78.98°F), and the ambient wind was stationary (0.0 m/s) at all participants' times. This study was approved by our local Ethics Review Board(Approval number:22-27).

\subsection{Tracking System}
\begin{figure}[h!]
    \centering
    \includegraphics[keepaspectratio, width=1.0\textwidth]{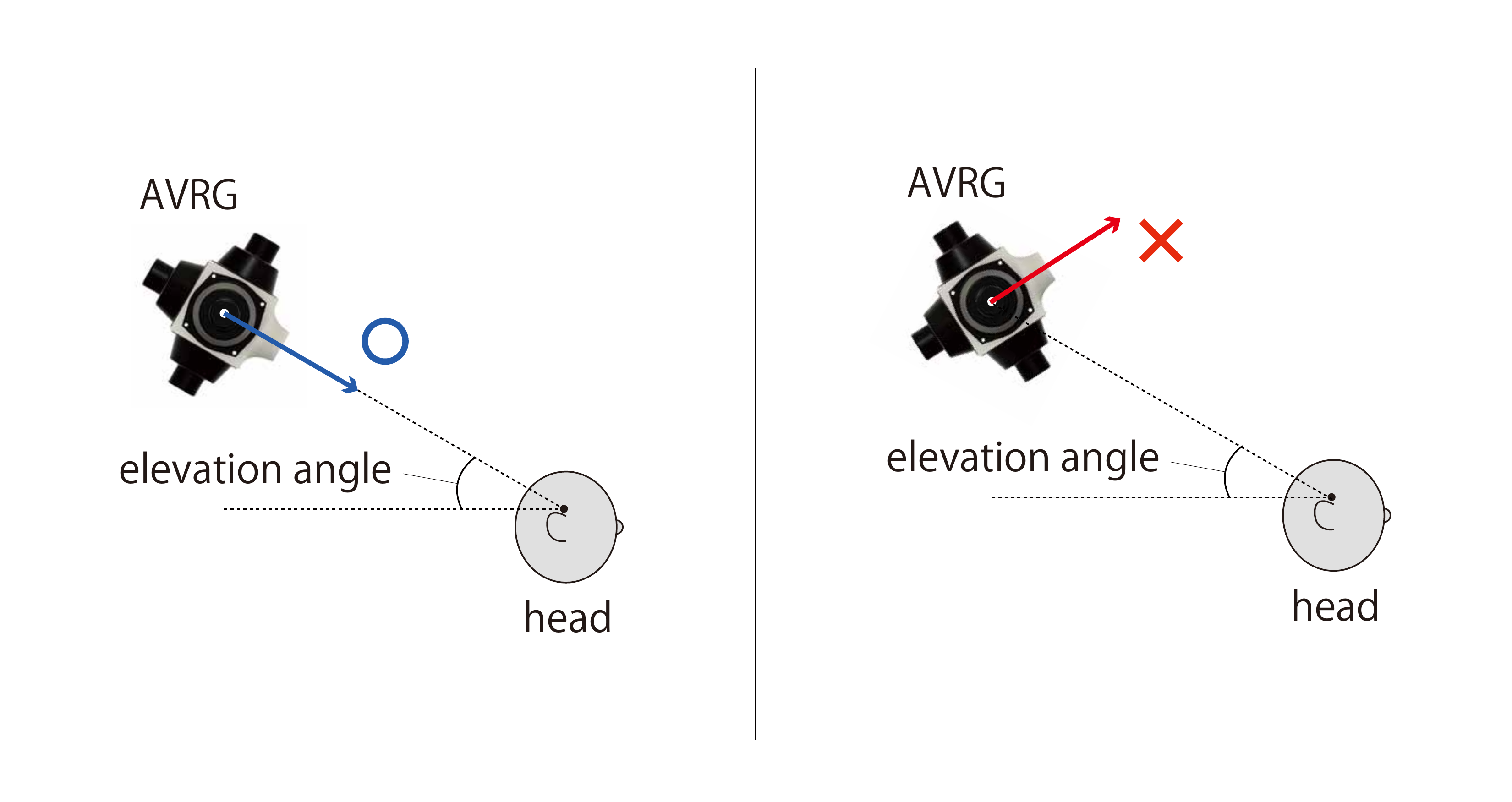}
    \caption{Directional alignment of the AVRG. Motion capture was used to determine not only whether the AVRG was on an elevation angle with the participant's head as the center coordinate but also whether the AVRG's ejection aperture was facing the participant's head.}
    \label{figure_user_study_angle_error}
  \end{figure}

 In these user studies, the distance between the AVRG and the participant and the presentation position was varied. Therefore, measuring the distance and angle between the head and the AVRG is necessary to hit the head with the air vortex ring. A tracking system was developed using motion capture (seven Prime17W, two PrimeX22, and one Prime41, OptiTrack). For the air vortex ring to hit the participant’s head, the participant’s head must be positioned on the mean vector from the opening of the AVRG. The coordinates of the center of gravity of the two points and the Euler angle of the AVRG were used to calculate the displacement between the normal vector of the aperture and the participant’s head, as shown in Fig.~\ref{figure_user_study_angle_error}. The allowable error in the position of the center of gravity of the AVRG was set at 50 mm.
 
\subsection{Calibration}
 In these user studies, it was necessary to identify the center of gravity of the AVRG and the center of gravity of the participant's head and to measure the distance and angle between them. However, it was not easy to perform the measurements with retroreflective markers on the participants' heads because it was anticipated that holding back their hair would affect their perception of the air vortex ring. The retroreflective markers were attached to the AVRG, and the distance and angle between the center of gravity of the AVRG and the position of the markers on the height-adjustable tripod were measured. The distance and angle between the center of gravity of the AVRG and the marker on the tripod were measured, and the position of the AVRG was determined by marking it on the floor.

 \subsection{Participants} 
 We recruited five DHHs (hearing level 90-130 dB, mean age 35.8 years old) in User Study 1, as shown in Table~\ref{table_participates_1}. Additionally, we recruited nine DHHs (hearing level 90-130 dB, mean age 29.1 years old) in User Study 2, as shown in Table~\ref{table_participates_2}, but four of User Study 2’s participants continued to participate in User Study 1. Participants were asked not to wear their hearing aids and cochlear implants during the experiment. Because of the high hearing levels of the participants in this study, they were not aware of the sound generated by the AVRG. Participants were entitled to receive compensation of 860 JPY. None of the participants had experienced the haptic sensation of the AVRG before.

\begin{table}[h!]
    \centering
    \caption{Participants' Information for User Study 1 (hearing level 90-130 dB, mean age 35.8 years old).}
    \includegraphics[keepaspectratio, width=1.0\textwidth]{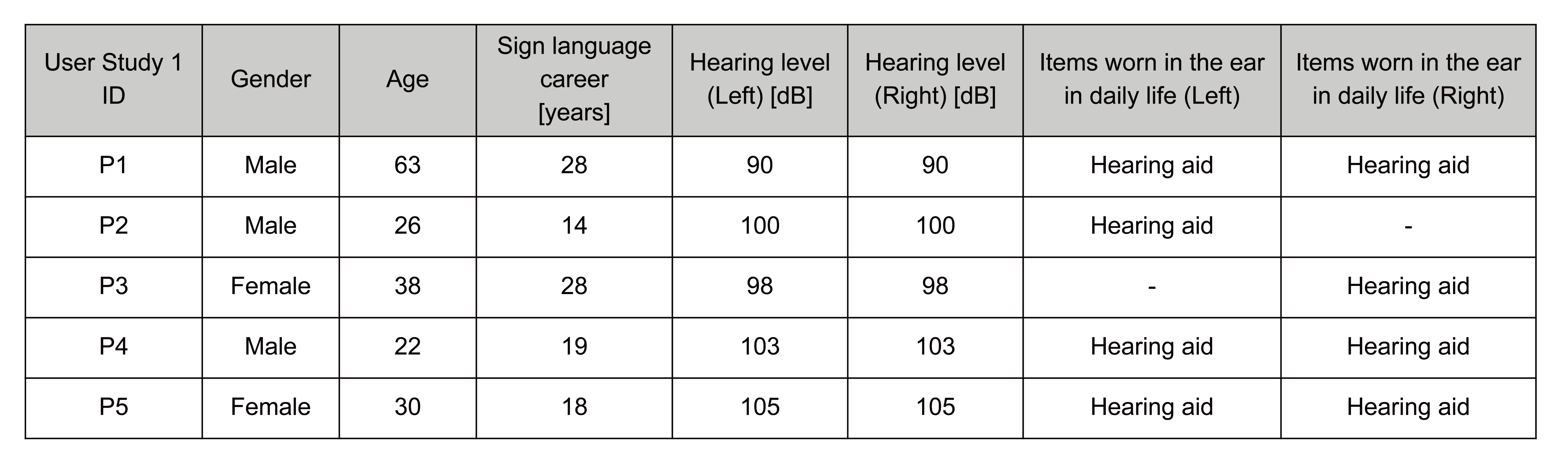}
    \label{table_participates_1}
\end{table}

\begin{table}[h!]
    \centering
    \caption{Participants' Information for User Study 2 (hearing level 90-130 dB, mean age 29.1 years old).}
    \includegraphics[keepaspectratio, width=1.0\textwidth]{figures/Table2.pdf}
    \label{table_participates_2}
\end{table}
\section{User Study 1}
\begin{figure}[h!]
    \centering
    \includegraphics[keepaspectratio, width=1.0\textwidth]{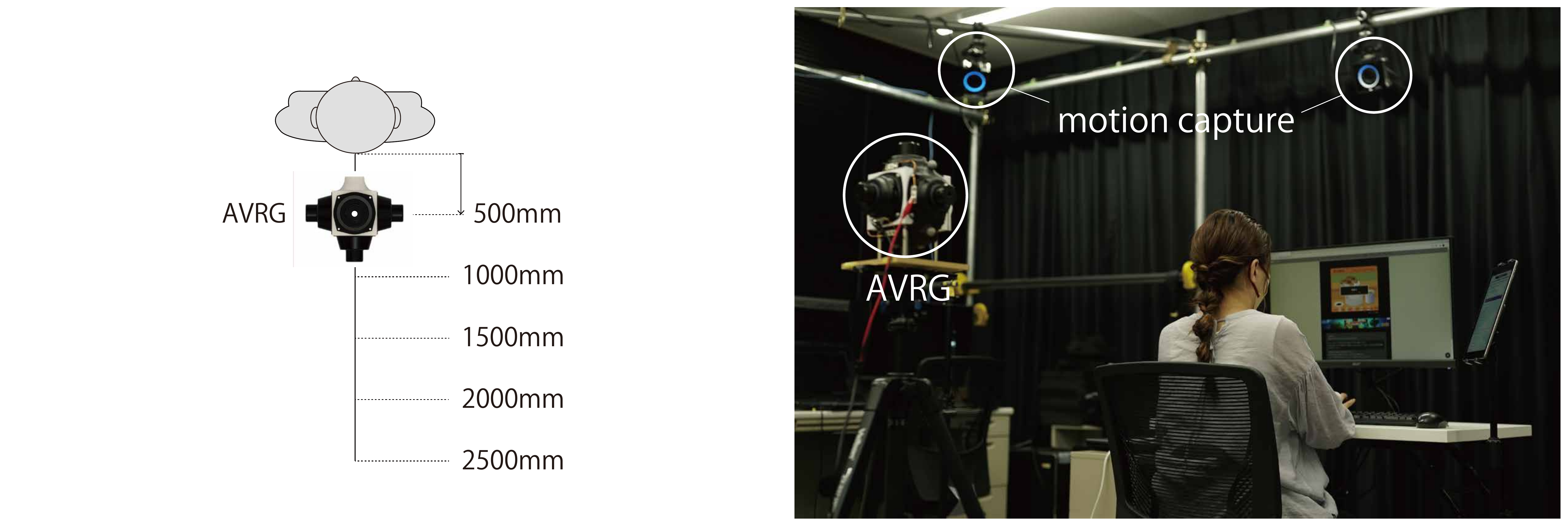}
    \caption{Overview of User Study 1. User Study 1 was conducted under five conditions with AVRG and the participant's head distances of 500, 1000, 1500, 2000, and 2500mm.}
    \label{figure_distance_experiment}
 \end{figure}

As shown in Fig.~\ref{figure_distance_experiment}, we conducted User Study 1 to determine how the perception of DHHs changes as the ejection distance and amount of ejected motion change. Here we investigated whether AVRGs could make us aware of DHH, how long it took us to notice them, how easy they were to notice qualitatively, and how comfortable they felt.

\subsection{Task and Procedure}
 In User Study 1, participants were tested on all combinations (25 trials) of the following parameters.

 \begin{itemize}
    \item Distances between AVRG and the participant's head\\
    ($distance = 500,1000,1500,2000,2500mm$).
    \item Roundness coefficients of speaker waveform ($b =0.001,0.002,0.003,0.004,1$).
 \end{itemize}

 After 15 trials, a 5-minute break was taken, and 10 trials were performed. Participants were asked to straighten their backs and perform the typing task without moving their heads. During the task, one air vortex ring was ejected every second, hitting the back of the participant's head a total of three times, and the participant was asked to press a button when he/she noticed that the air vortex ring had hit his/her head. Then we asked, ``Was this wind easy to notice?'' and ``Do you think this wind is comfortable?'' on a 7-point Likert scale (1 strongly disagree to 7 strongly agree).

 \subsection{Result} 
 \begin{figure}[h!]
    \centering
    \subfigure{\includegraphics[width=0.48\textwidth]{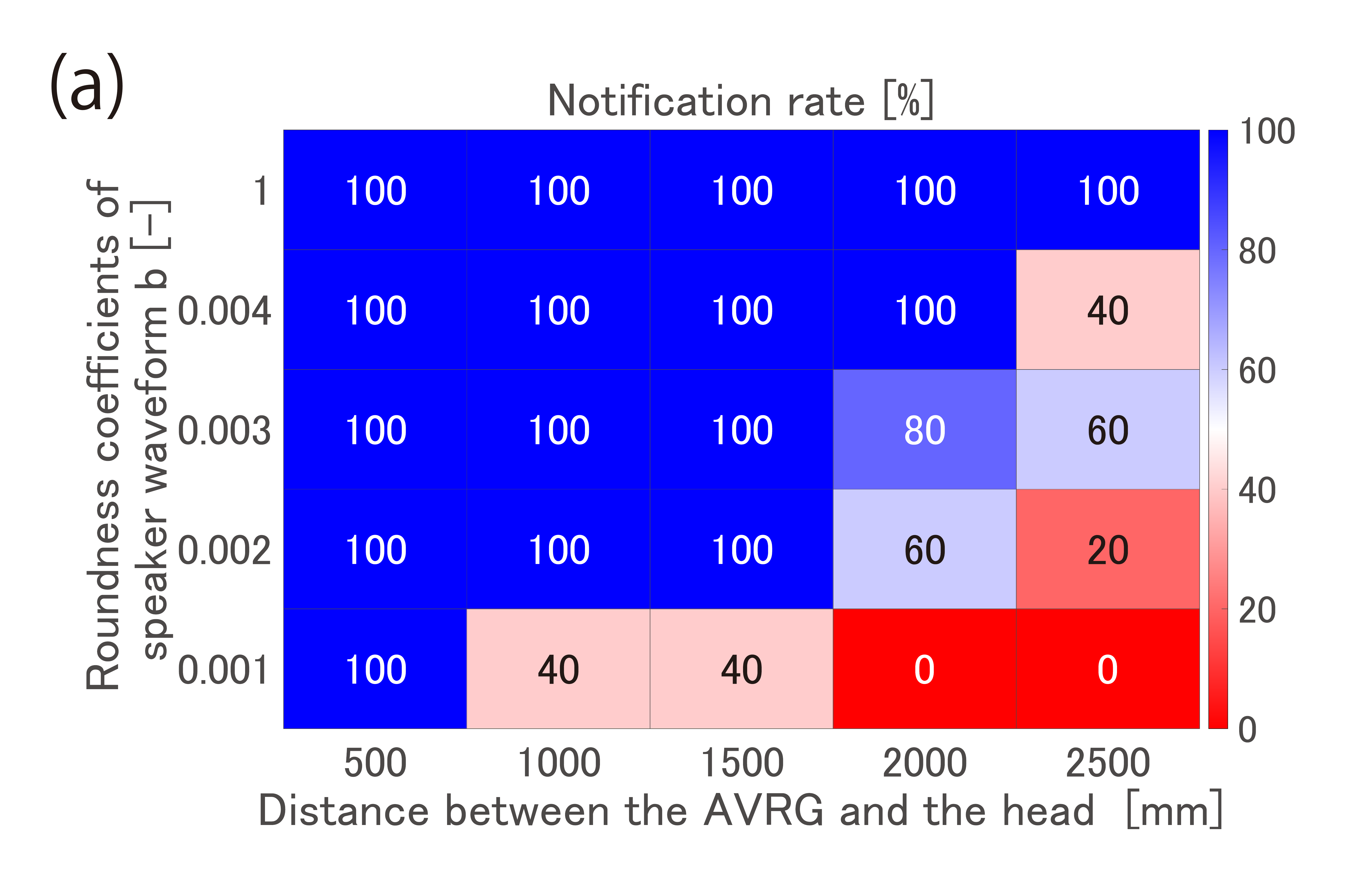}}
    \subfigure{\includegraphics[width=0.48\textwidth]{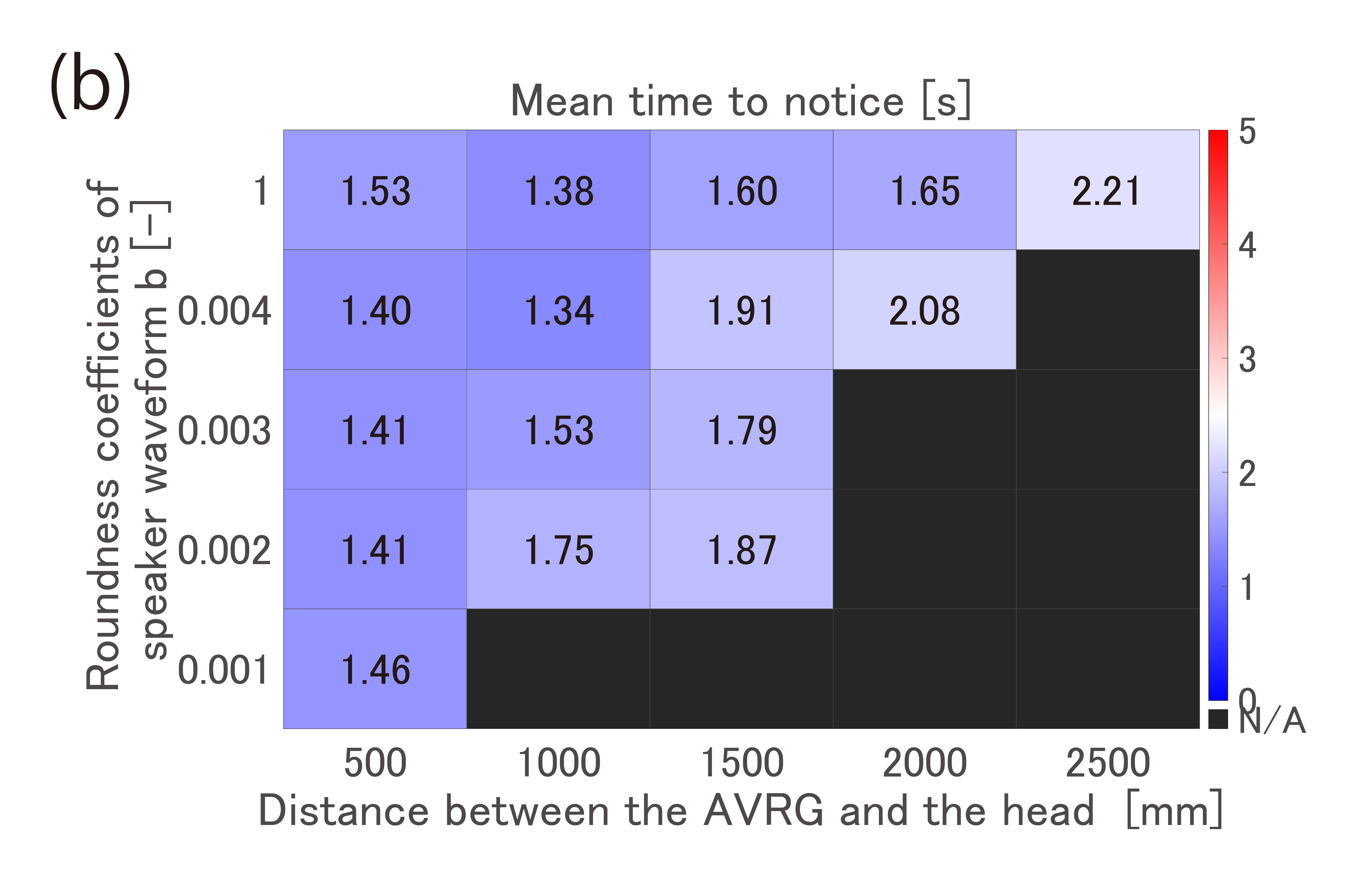}}
    \subfigure{\includegraphics[width=0.48\textwidth]{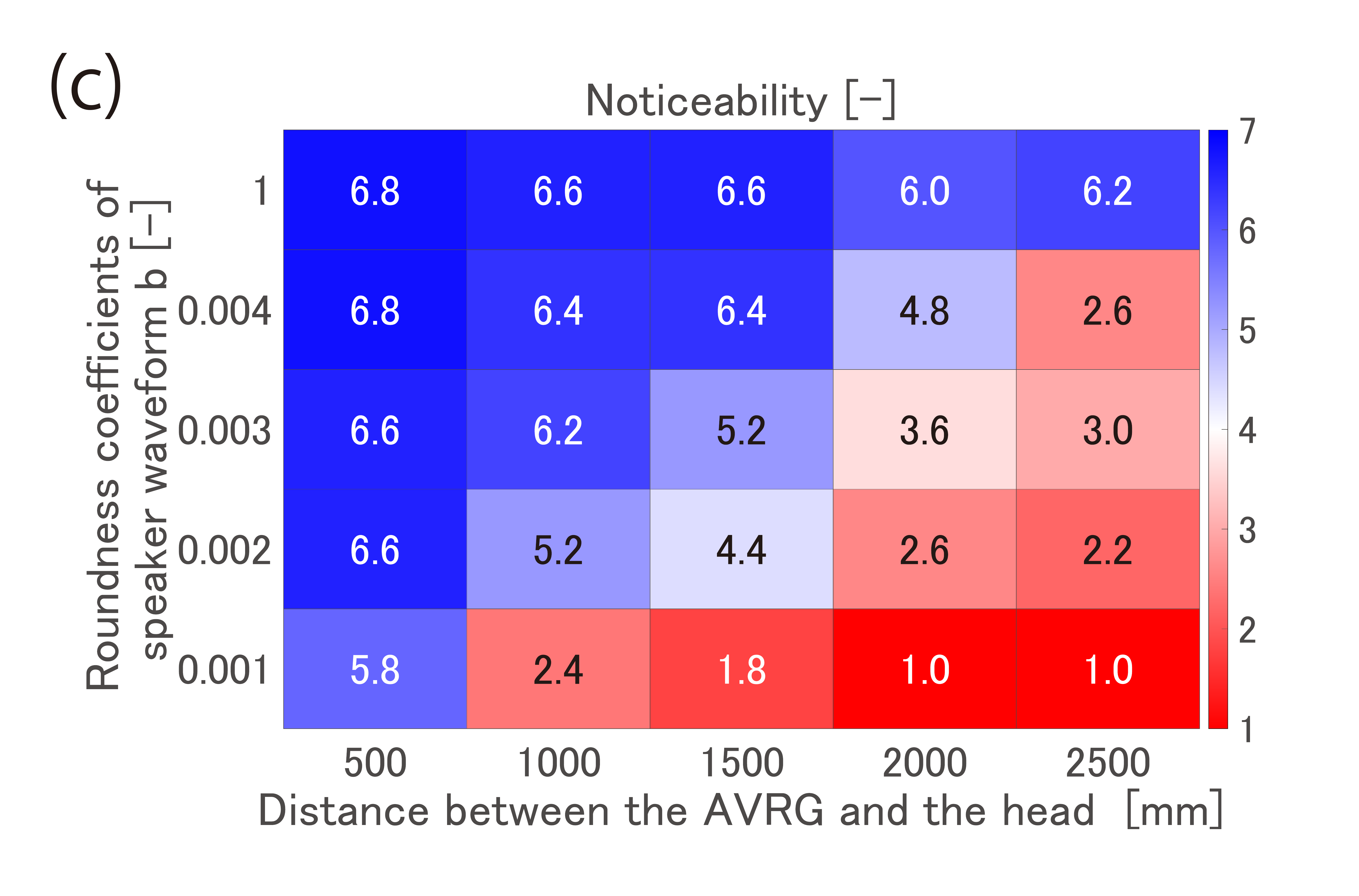}}
    \subfigure{\includegraphics[width=0.48\textwidth]{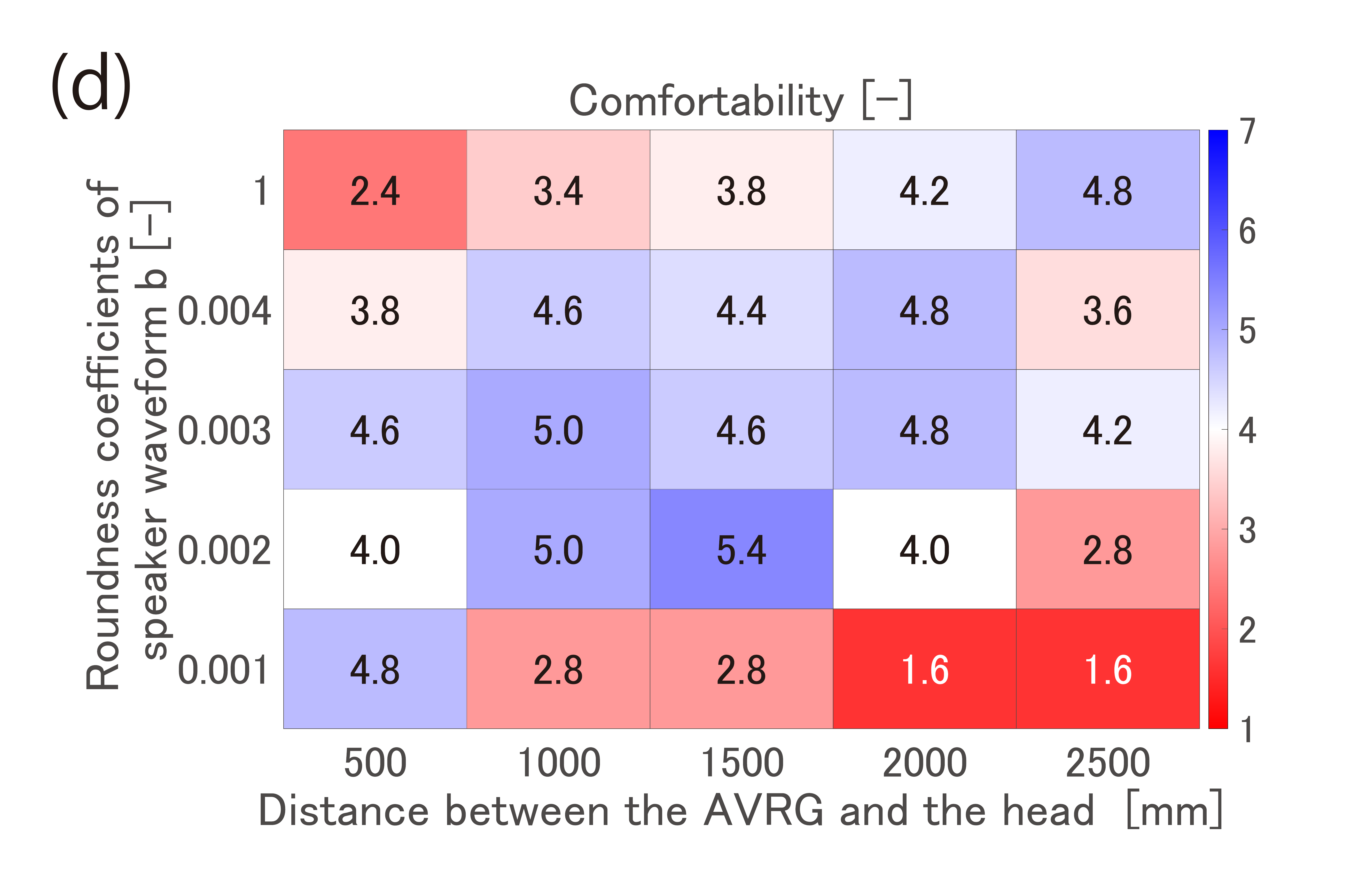}}
    \caption{Results of User Study 1. (a) Notification rate [\%]. The closer the speaker's roundness coefficient is to one, increasing the distance the participant can notice. (b) Mean time from the time the air vortex ring was emitted to the time to notice [s]. The condition with the shortest mean time to notice was at a distance of 1000 mm and roundness coefficients of 0.004. (c) Noticeability [-]. Overall, the greater the distance and the smaller the roundness coefficients, the smaller the noticeability.
    (d) Comfortability [-]. The comfortability of the air vortex ring was highest when the air vortex was barely noticeable, as shown in Fig.~\ref{figure_userstudy1} (a) and  Fig.~\ref{figure_userstudy1} (c).}
    \label{figure_userstudy1}
 \end{figure}
 
  Fig.~\ref{figure_userstudy1} (a)(b) shows the quantitative results of User Study 1. Fig.~\ref{figure_userstudy1} (a) shows the notification rate, indicating whether the participant could notice and push the button within 5 seconds after the AVRG emitted an air vortex ring. 
All participants noticed the haptic sensation of the air vortex ring when the distance was within 500mm, 1500mm, 1500mm, 2000mm, and 2500mm for the roundness coefficients b of 0.001, 0.002, 0.003, 0.004, and 1, respectively.
The closer the speaker's roundness coefficient is to one, the faster the piston velocity and the greater the momentum that AVRG can generate, thus increasing the distance the participant can notice.
When compared to Fig.~\ref{figure_result_of_fur_experiment}, the conditions under which the fur sway was detected are similar to those that were noticed by the participant.

  Fig.~\ref{figure_userstudy1} (b) shows the mean time from the time the air vortex ring was emitted to the time to notice; in the condition where one or more participants failed to notice the air vortex ring within 5 s, the mean time was set to N/A because it was impossible to obtain an accurate mean time. The condition with the shortest mean time to notice was at a distance of 1000 mm and an acceleration coefficient of 0.004. The mean time to notice tended to increase as the distance increased. The highest mean time was 2.21 s in the condition in which 100\% of the time was taken to notice.
 
Fig.~\ref{figure_userstudy1} (c) shows the mean value of the seven-level evaluation of noticeability. 
Overall, the greater the distance and the smaller the roundness coefficients, the smaller is the noticeability.
As shown in Fig.~\ref{figure_speaker_roundness} (b) and Fig.~\ref{figure_userstudy1} (c), the distance at which the participants could notice the air vortex ring increased as the roundness coefficients of the speaker approached 1. From Fig.~\ref{figure_userstudy1}(a) showing the results of quantitative evaluation, it was found that the notification rate was 100\% when the distance was 1000 mm and the roundness coefficient was 0.002, when the distance was 1500 mm and the roundness coefficient was 0.003, and when the distance was 2000 mm and the roundness coefficient was 0.004, respectively. However, in Fig.~\ref{figure_userstudy1}(c), the result of the qualitative evaluation shows that the noticeability was 4.8-5.2 in these cases, indicating that the noticeability was lower for conditions with larger roundness coefficients and shorter distances.

  Fig.~\ref{figure_userstudy1} (d) shows the mean value of the seven-level evaluation of comfortability. The highest noticeability is found when the distance is 500 mm and the tame coefficient is 1, while the comfortability is 2.4, the third lowest value among all 25 conditions. The comfortability of the air vortex ring was highest when the air vortex was barely noticeable, as shown in Fig.~\ref{figure_userstudy1} (a) and  Fig.~\ref{figure_userstudy1} (c). 

 \subsection{Discussion}
 The results in Fig.~\ref{figure_userstudy1} (a) and  Fig.~\ref{figure_userstudy1} (c) show that if the roundness coefficient is 1, there is a 100\% notification rate and 6.2/7 noticeability even when the distance to the stationary target DHH is approximately 2500mm away. Fig.~\ref{figure_result_of_fur_experiment} shows the gap between the target and the hit position is about 100mm, making it difficult to hit the desired position in the head. Still, it may be possible to notify the stationary DHH as an initial conversation cue.

 However, the mean time to notice in Fig.~\ref{figure_userstudy1} (b) indicates that the time from injection to recognition is approximately 0.72 m/s, which is 470 times slower than the speed of sound (340 m/s), such as voice or notification sounds. Therefore, we believe that the air vortex ring speed and the haptic sense generate limitations. It is necessary to verify the comfortability of the talker using AVRG as future research and to discuss further whether it can be used as an application.
 
 The comfortability in Fig.~\ref{figure_userstudy1} (d) shows optimal roundness coefficients for each distance from the target DHH to make the target feel comfortable. Based on this, we believe a notification to the DHH can be sent without irritating them by first measuring the distance to the target and projecting air vortex rings with the speaker waveform with the most comfortable roundness coefficients.

\section{User Study 2}
 We conducted User Study 2 to investigate how the perception of DHHs changes when the direction of the air vortex ring ejection relative to the head is changed. Specifically, we investigated the direction from which people would notice the air vortex ring when it was applied from a specific direction. We also investigated qualitatively how easy it was to notice and how comfortable it felt as an initiating conversation.

\begin{figure}[t!]
    \centering
    \subfigure{\includegraphics[keepaspectratio, width=1.0\textwidth]{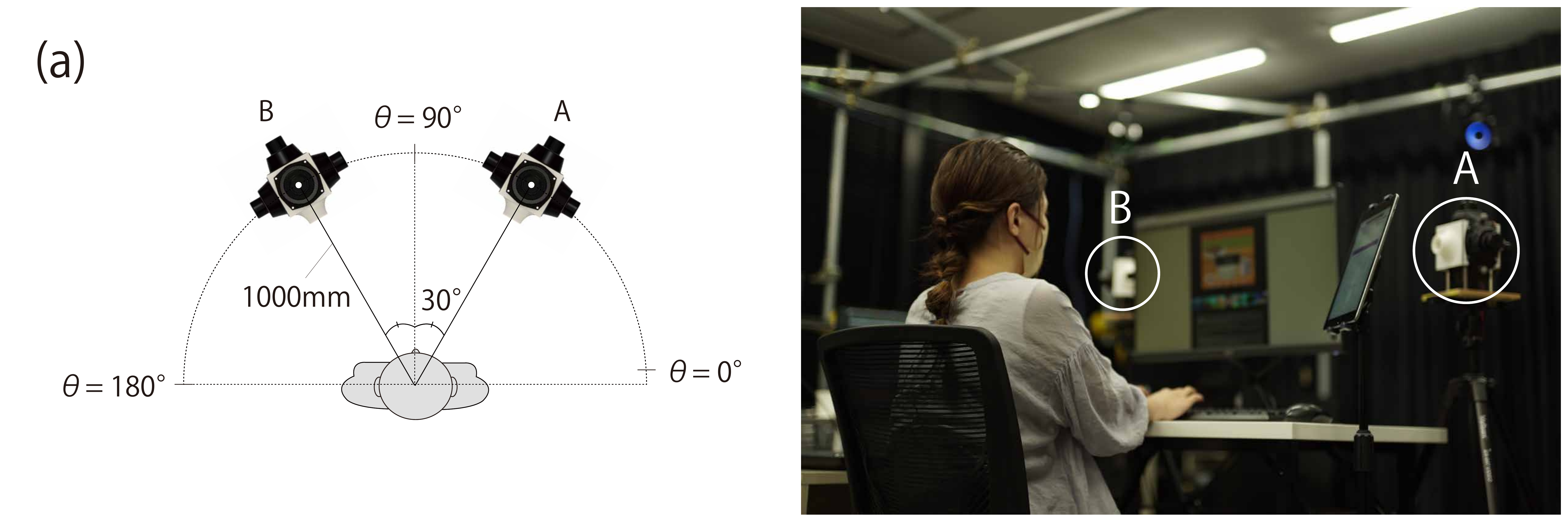}}
    \subfigure{\includegraphics[keepaspectratio, width=1.0\textwidth]{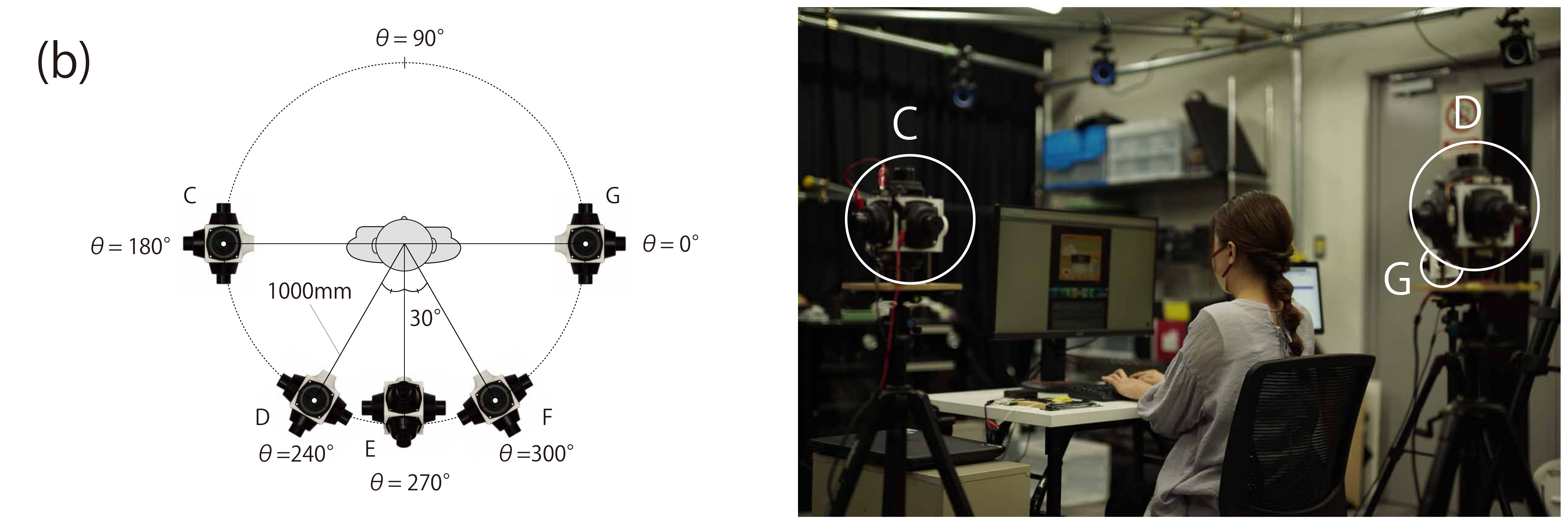}}
    \subfigure{\includegraphics[keepaspectratio, width=0.5\textwidth]{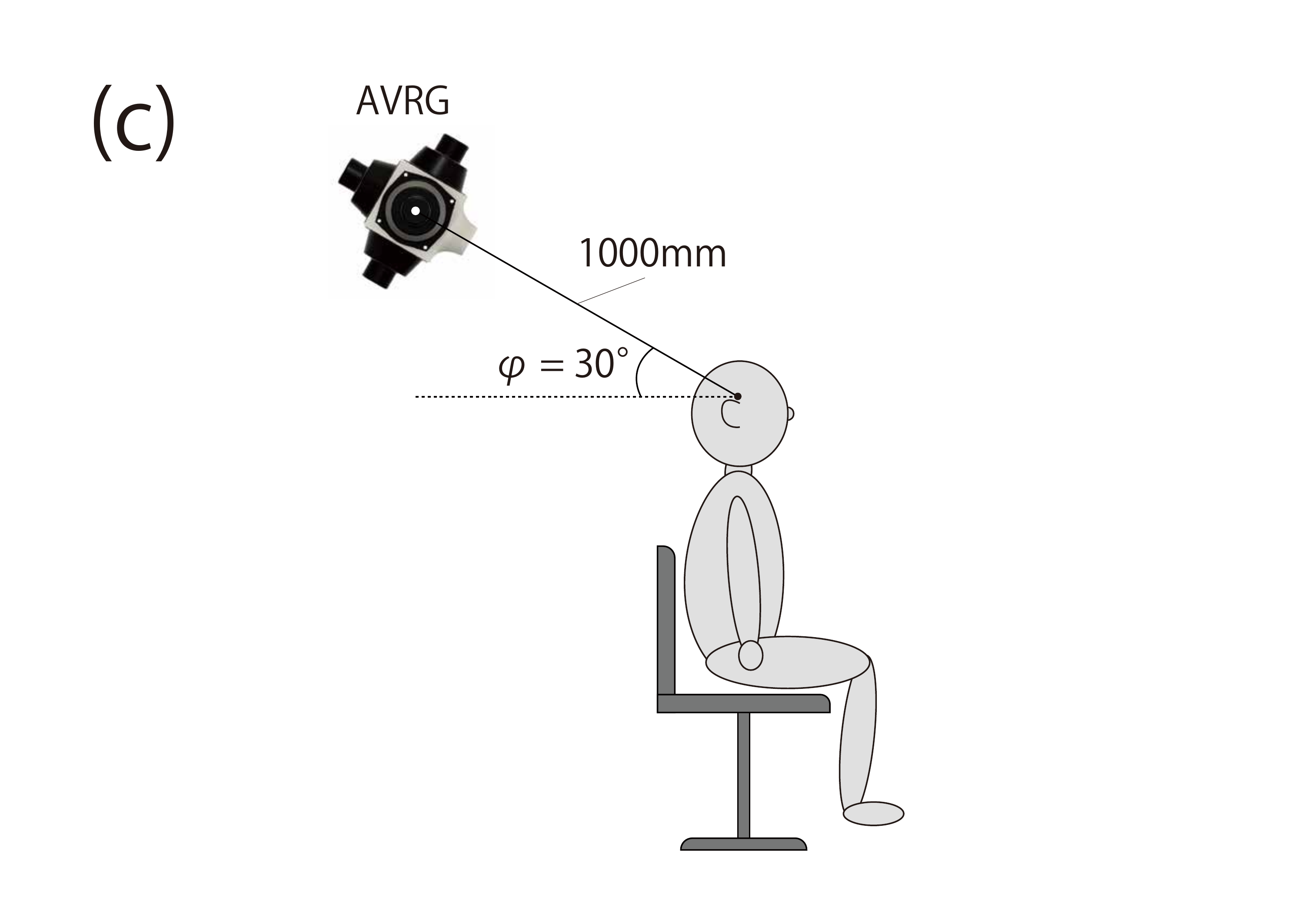}}
    \caption{Overview of user study 2.
    (a) Experiment to hit air vortex ring from forward directions, A:($\theta$ = 60°, $\varphi$ = 0°) , B:($\theta$ = 120°, $\varphi$ = 0°).
    (b) Experiment to hit an air vortex ring from lateral and backward directions, C:($\theta$ = 180°, $\varphi$ = 0°), D:($\theta$ = 240°, $\varphi$ = 0°), E:($\theta$ = 270°, $\varphi$ = 30°), F:($\theta$ = 300°, $\varphi$ = 0°), and G:($\theta$ = 360°, $\varphi$ = 0°).
    (c) Side view of the experiment in which an air vortex ring was hit from the backward direction E:($\theta$ = 270°, $\varphi$ = 30°.)}
    \label{figure_user_study_2}
 \end{figure}

 \begin{figure}[t!]
    \centering
    \includegraphics[keepaspectratio, width=1.0\textwidth]{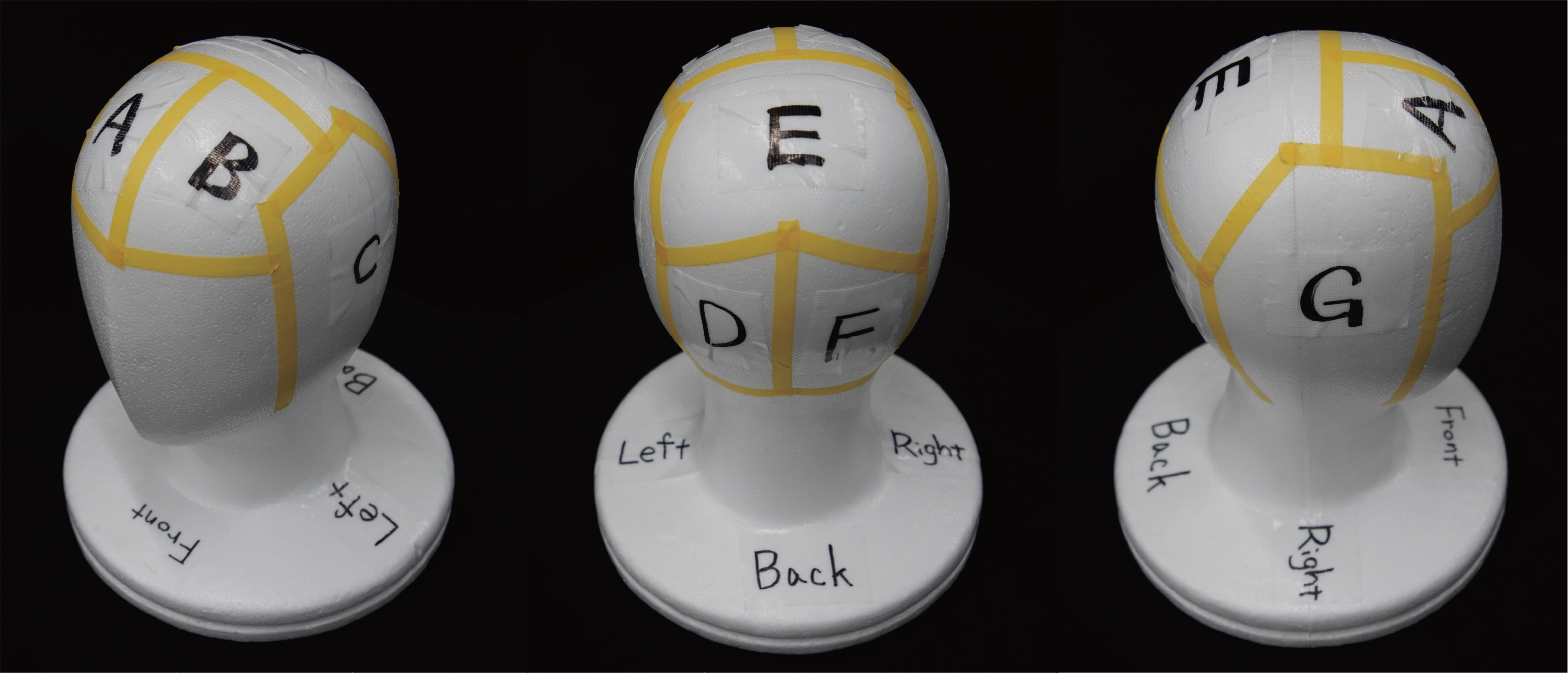}
    \caption{Mannequin head to point to the position hit by the air vortex ring corresponding to the 7 directions in Fig.~\ref{figure_user_study_2}(a)(b).}
    \label{figure_head}
 \end{figure}
 
 \subsection{Task and Procedure}
 In User Study 2, the distance was fixed at 1000mm. The speaker's roundness coefficients were fixed at b=1, and the speaker displacement was fixed at 8.2mm. We conducted the experiment under these conditions to determine whether we could hit the desired location without inaccuracy and whether the participant noticed the air vortex ring. As shown in Fig.~\ref{figure_user_study_2} (a)(b)(c), the participant's horizontal angle is represented by $\theta$, and the elevation angle is represented by $\varphi$. First, an experiment was conducted in which air vortex rings were applied from two directions, 
 \begin{itemize}
    \item A: Right frontal head ($\theta$ = 60°, $\varphi$ = 0°),
    \item B: Left frontal head ($\theta$ = 120°, $\varphi$ = 0°),
 \end{itemize}
 as shown in Fig.~\ref{figure_user_study_2} (a). A total of three trials (two directions and one of the two directions) were performed randomly.

 In addition, as shown in Fig.~\ref{figure_user_study_2} (b), experiments were conducted in which the participants were hit from five different directions:
 \begin{itemize}
    \item C: Left temporal ($\theta$ = 180°, $\varphi$ = 0°),
    \item D: Left occipital ($\theta$ = 240°, $\varphi$ = 0°),
    \item E: Parietal ($\theta$ = 270°, $\varphi$ = 30°),
    \item F: Right occipital ($\theta$ = 300°, $\varphi$ = 0°),
    \item G: Right temporal ($\theta$ = 360°, $\varphi$ = 0°).
 \end{itemize}
 A total of six trials (five directions and one of the five directions) were performed randomly.

Participants were instructed not to move their heads during the experiment. In each condition, the participants performed the typing task as in User Study 1, and when they noticed the air vortex ring hit their head, they pressed the button to measure the time to notice. The participants were then asked to respond to the area of the head that was hit using the mannequin head shown in Fig.~\ref{figure_head} and were asked, ``Did you find this wind easy to notice?'' ``Did you think it was comfortable to talk to?'' on a seven-point scale (1 strongly disagree to 7 strongly agree).

\subsection{Result}
 \begin{figure}[h!]
    \centering
    \subfigure{\includegraphics[width=1.0\textwidth]{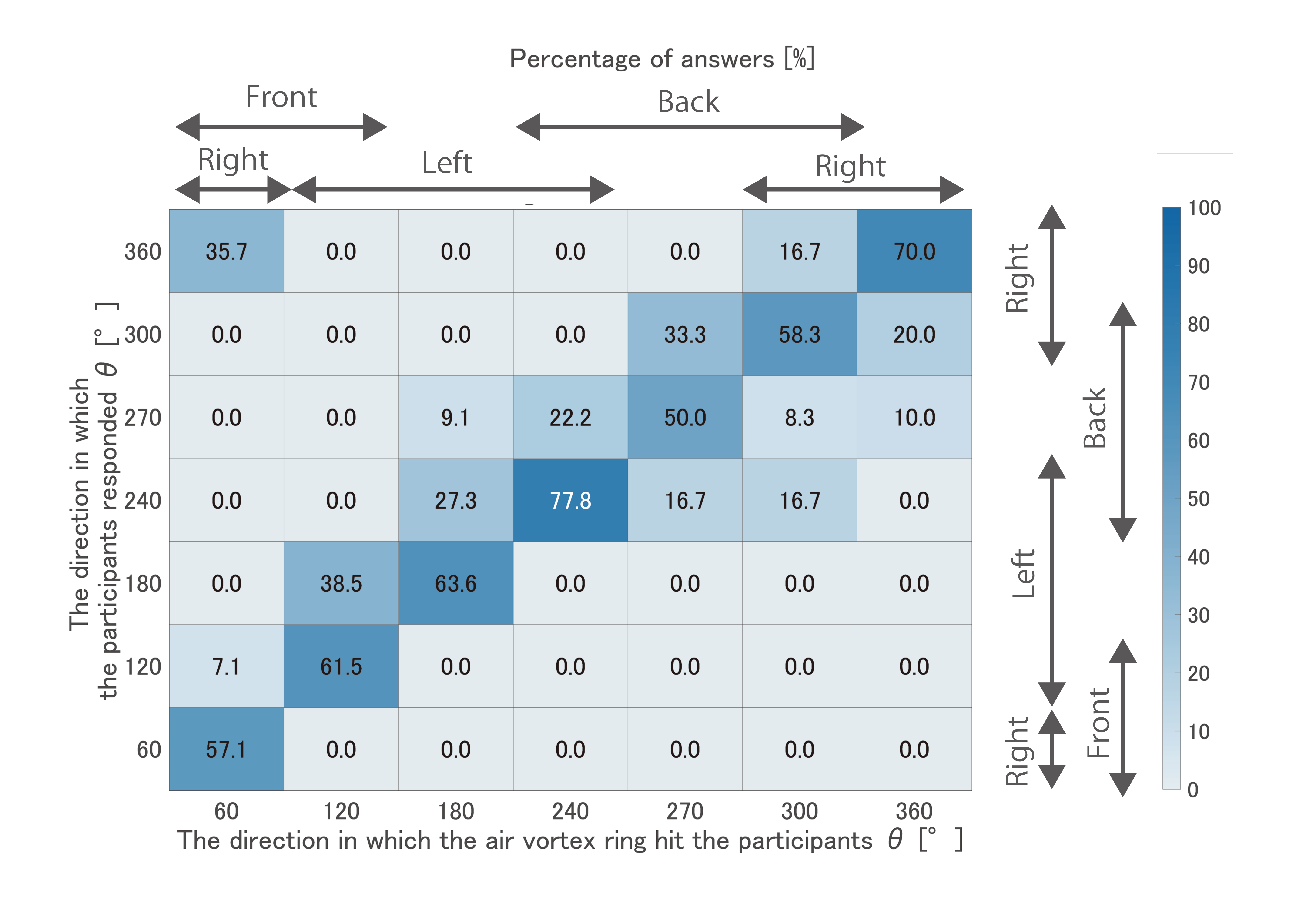}}  
    \caption{Percentage of responses when the air vortex ring is hit from seven directions in Fig.~\ref{figure_user_study_2} [\%].}
    \label{figure_button_userstudy2}
 \end{figure}
 
 Fig.~\ref{figure_button_userstudy2} shows the percentages of responses in User Study 2, in which the participants were hit from seven directions. The percentage of responses in the same direction as the direction in which the air vortex ring was hit was 57.1\% for $\theta$ =  60°, 61.5\% for $\theta$ =  120°, 63.6\% for $\theta$ = 180°, 77.8\% for $\theta$ = 240°, 50.0\% for $\theta$ = 270°, 50.0\% for $\theta$ = 300°, 58.3\% for $\theta$ = 270°, and 70.0\% for $\theta$ = 360°. 

When the air vortex rings were hit from the right side of the participants ($\theta$ =  60°, $\theta$ = 300°, $\theta$ = 360°), the percentage of responses that hit from the right side was 92.9\%, 75\%, and 90\%, respectively. When the air vortex rings were hit from the left side of the participants ($\theta$ =  120°, $\theta$ = 180°, $\theta$ = 240°), the percentage of responses that hit from the left side was 100\%, 90.9\%, and 100\%, respectively.

 \begin{figure}[h!]
    \centering
    \subfigure{\includegraphics[width=0.87\textwidth]{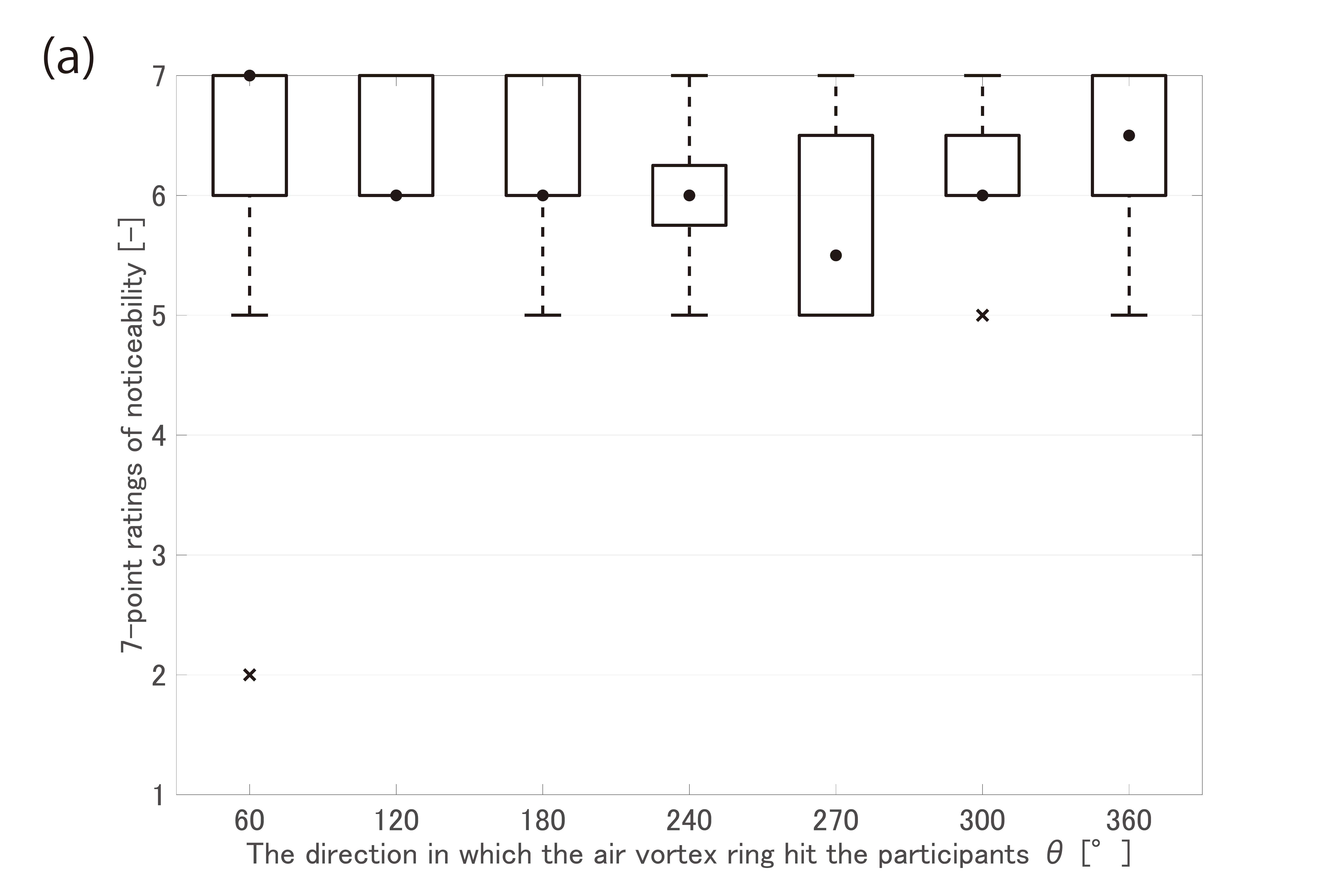}}
    \subfigure{\includegraphics[width=0.87\textwidth]{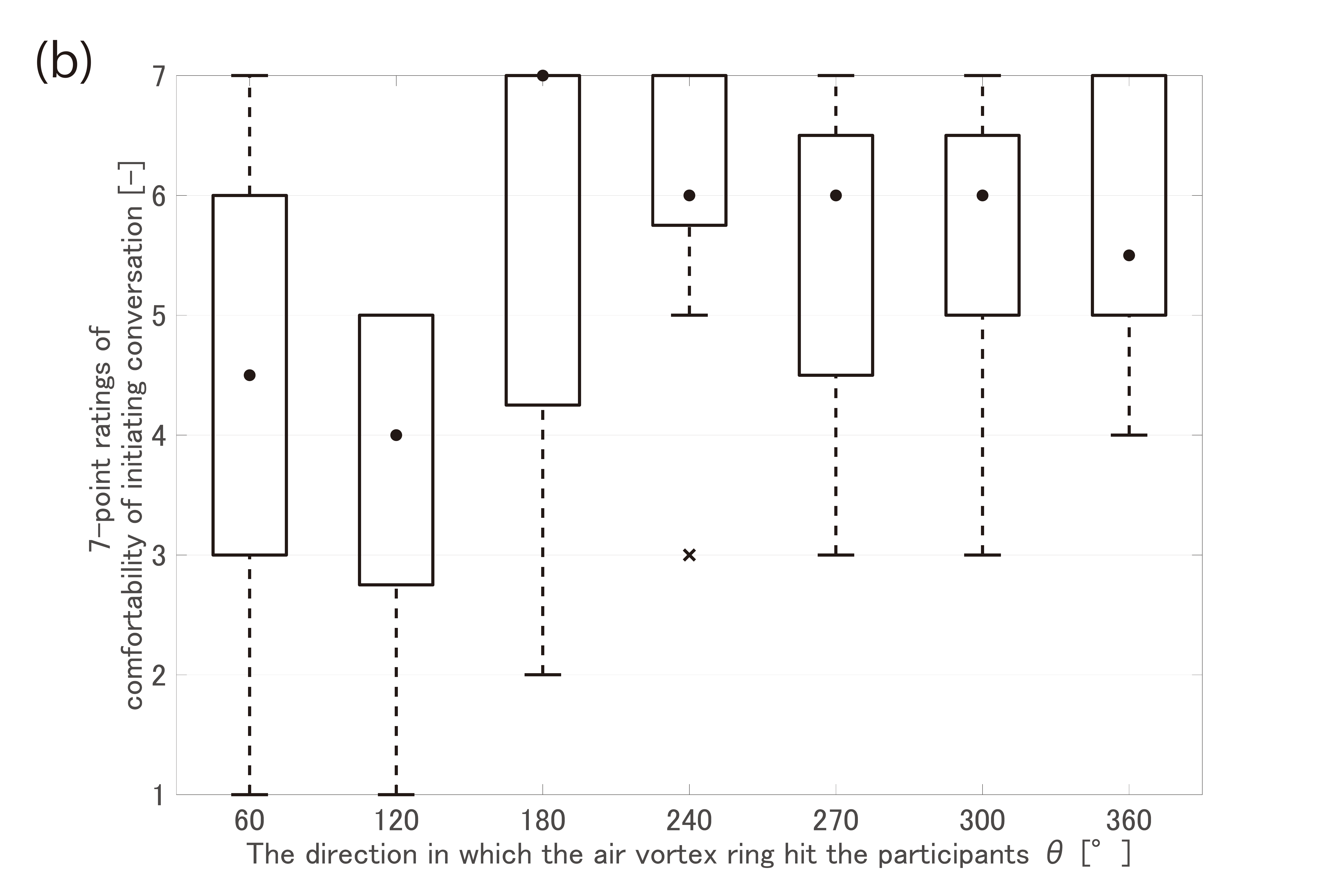}} 
    \caption{7-point ratings of the subjective parameters for seven directions. The median is represented by a dot in the middle of the box on each boxplot. The 25th and 75th percentiles are represented by the bottom and top edges of the box, respectively. Data points that are considered outliers are shown as individual 'x' markers, and the whiskers extend to the most extreme data points that are not considered outliers. (a) noticeability [-]. (b) comfortability [-].}
    \label{figure_Qualitative_userstudy2}
 \end{figure}
 
 \newpage
 Fig.~\ref{figure_Qualitative_userstudy2} shows the boxplots of the 7-point ratings of the subjective parameters of noticeability and comfortability of initiating the conversation for each direction. The median is represented by a dot in the middle of the box on each boxplot. The 25th and 75th percentiles are represented by the bottom and top edges of the box, respectively. Data points that are considered outliers are shown as individual 'x' markers, and the whiskers extend to the most extreme data points that are not considered outliers.

In Fig.~\ref{figure_Qualitative_userstudy2}(a), which represents the noticeability, the median value for all directions except direction $\theta$ = 270° was greater than 6, while only direction $\theta$ = 270° was difficult to notice, with the median value of 5.5. Additionally, the 75th percentiles were never less than 5 in all directions. In Fig.~\ref{figure_Qualitative_userstudy2}(b), which represents the comfortability, the median value was approximately 1 to 3 lower in the forward condition than in the backward conditions. The median value was 5.5 or higher in the backward condition, indicating that the subject was relatively comfortable.

 \subsection{Discussion}
 In the discussion on whether AVRG can be used to indicate the direction of speech, it was found that more than 50\% of the respondents noticed the same direction as the actual direction, and more than 75\% of the respondents were able to determine whether the direction was left or right. It is thought that AVRG can be used to indicate the direction in which the user is spoken to.

Although the results also showed that it was qualitatively easy to notice all directions, there is much concern about whether the air vortex ring will hit the eyes in the forward direction as a comfort feature of speech, and it is necessary to improve the system so that the air vortex ring can be applied with higher accuracy. In the back of the head, there are a few areas where the air vortex ring hits and offends the user, so the current device is considered to be highly effective.

\section{Discussion and Application Design}
\subsection{Qualitative feedback}
\begin{table}[h!]
    \centering
    \caption{Is there anything you usually do to notice that you are being spoken to?}
    \begin{tabular}{|L{14cm}{}{m}|}
        \hline
        Original Answers
        \nextRow
        \hline \hline
        Stand where I can notice visually and wear a hearing aid.
        
        When I am concentrating on typing, etc., I am startled if someone taps me on the shoulder, so I move away from the screen as much as possible so that I can also see the side of the screen (keep my arms outstretched).
        
        Look the other person in the eye.
        
        I listen with my ears as much as possible by utilizing my cochlear implant.
        
        When someone is about to talk to me, I pay attention 360 degrees (by glancing at them, etc.)
        
        I am aware of the presence of others.
        
        I don't usually do anything in particular. In unfamiliar places, I look around nervously.
        \nextRow
        \hline
    \end{tabular}
    \label{table_talk_daily}
\end{table}

\begin{table}[h!]
    \centering
    \caption{Please describe any good or bad points when assuming that you will be spoken to by the AVRG system.}
    \begin{tabular}{|L{14cm}{}{m}|}
        \hline
        Original Answers
        \nextRow
        \hline \hline
        It feels unnatural and crummy. (It is different from just wind, so it could be said that it is easy to understand, but it is somewhat similar to the feeling of being blown by a hit of air, which may not be a good feeling)
        
        When I feel the wind from behind me, especially if it's a pinpoint wind, I wonder if it's huffing and hiting from someone else's mouth. I get scared.
        
        From the front or side, there was a little resistance when it hit my eyes, but from behind, there was no resistance whatsoever.
        
        Since it is the wind, it is good that the person speaking to me does not have to worry about the force with which he or she taps me on the shoulder. The person being called is also less surprised by the wind. Ideally, it would be better to have a forehead as well because if it is near the eyes, the person would be surprised by touching the eyeballs.
        \nextRow
        \hline 
    \end{tabular}
    \label{table_goodbad_point}
\end{table}

\begin{table}[h!]
    \centering
    \caption{Please tell us what needs to be improved for future development as an assistive device for talking to DHH.}
    \begin{tabular}{|L{14cm}{}{m}|}
        \hline
        Original Answers
        \nextRow
        \hline \hline
        I thought the AVRG injection interval could be a little shorter.
        
        I think it is interesting in terms of using air. I think it would be easier to understand if there is a mechanism to understand the rhythmic changes.
        
        I thought the strength of the wind was okay, but a little louder than one point would be better.
        
        Is it a disadvantage that this system requires the equipment to be fixed? If it floats like a drone, I would appreciate more usage scenarios. Realistically, I think it would be like a small electric fan worn around the neck. 
        \nextRow
        \hline
    \end{tabular}
    \label{table_improvement}
\end{table}

We asked participants for feedback at the end of the user study about being spoken to and about the AVRG system. Table~\ref{table_talk_daily} shows the answers to the question, "Is there anything you usually do to notice that you are being spoken to?" Most of the participants answered that they try to look around with their eyes on a daily basis to notice what is being said. It is thought that the participants may not be able to concentrate deeply because they are using their visual field too much to compensate for various information. In addition, it is thought that AVRG can contribute to reducing this burden.

In Table~\ref{table_goodbad_point}, we summarized the good points and bad points of the system when it was assumed that the user would be spoken to.
The opinions were divided into two groups: those who said they were surprised by the unnaturalness of the air vortex ring, and those who said they were less surprised by the air pressure and felt no resistance compared to a tap on the shoulder.
During the experiment, some participants were surprised every time they felt the air vortex ring, while others gradually became accustomed to the tactile sensation of the air vortex ring. It may be necessary to adjust the force of the air vortex ring so that it is more optimized for each individual DHH.
Some participants also commented that they felt as if they were being poked in a single point against their heads. It is necessary to change the tactile sensation by adjusting the size of the air vortex ring rather than by adjusting the force alone.

Table~\ref{table_improvement} summarizes the points that must be improved in order for the device to be developed as an assistive device for talking to DHH in the future.
The respondents suggested that not only adjusting the force of the air vortex rings but also changing the rhythm of the air vortex rings would make the device easier to understand compared to wind. It was suggested that the environment be notified not only by air vortex rings flying from the ceiling of the room but also by a method in which the person being spoken to could hold a very small AVRG and speak to the listener from it.

\subsection{Limitation and Improvement}

\begin{figure}[h!]
    \centering
    \includegraphics[keepaspectratio, width=1.0\textwidth]{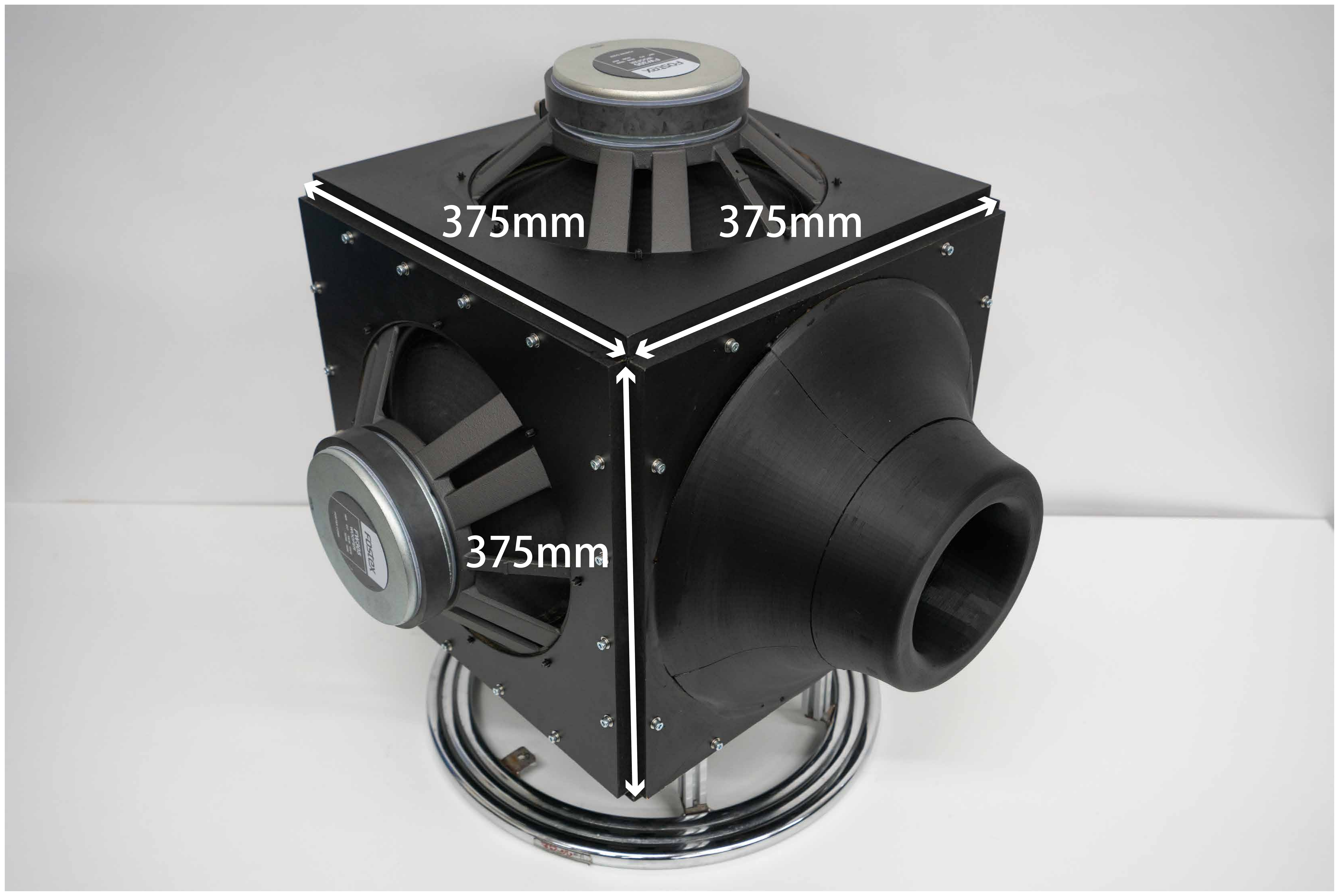}
    \caption{Huge AVRG. A nozzle with an aperture diameter of 120 mm and five speakers (FW305, Fostex) with a diameter of the speaker corn of 259 mm were mounted in an enclosure made of aluminum frame and MDF material.}
    \label{figure_Large_AVRG}
\end{figure}

As verified in the accuracy experiment, the problem is that the air vortex ring cannot hit any position on the head because of the large displacement of the ejection direction compared to the size of the human head, and it does not hit the head when the flying distance is 2.5 m or longer. One of the reasons for the lack of a straight flight is that the counter-rotating vortex at the aperture is created when the speaker membrane is pulled and interferes with the vortex when the air is pushed out, weakening it. Therefore, a method to make the waveform just push in without pulling the speaker membrane or a method to keep the speaker membrane in place after pulling it in and wait until the vortex diffuses can be considered. However, care must be taken to avoid a trade-off between the smaller displacement of the speaker and the smaller frequency of the air vortex rings.

Another limitation is that it is not possible to send the air vortex ring in an arbitrary direction or to aim it at a moving head. Therefore, it is necessary first to locate the head with a camera and then use a robot arm or motor to correct the direction of the AVRG toward the head. In addition, because it takes time for the air vortex ring to reach the head, it is necessary to calculate the injection angle by calculating backward from the time it takes to reach the head.

As another improvement, as shown in Fig.~\ref{figure_Large_AVRG}, a AVRG was fabricated which is larger than the AVRG shown in Fig.~\ref{figure_implementation_overview}. A nozzle with an aperture diameter of 120 mm and five speakers (FW305, Fostex) with a diameter of the speaker corn of 259 mm were mounted in an enclosure made of aluminum frame and MDF material. As shown in Equation~\ref{aperture_diameter}, the momentum of the air vortex ring generated by increasing the diameter of the speaker increases by the square of the diameter of the speaker. This can generate stronger pressure, which may be applicable to situations that are more difficult to notice. It can hit anywhere on the body of a person standing 5 meters away.

\begin{figure}[h!]
    \centering
    \includegraphics[keepaspectratio, width=1.0\textwidth]{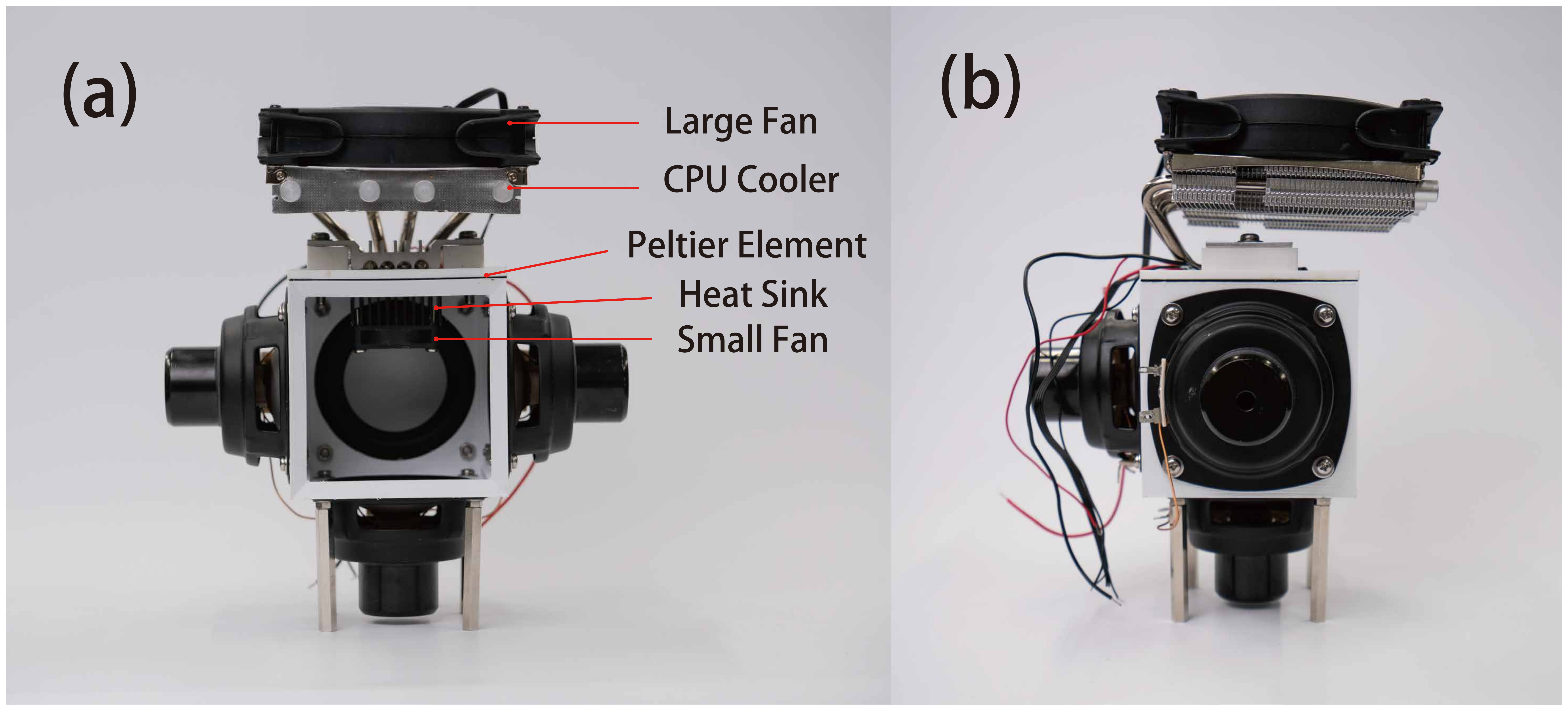}
    \caption{AVRG to create a temperature difference by the peltier element. A CPU cooler and a large fan are placed outside to release heat, and a heat sink and a small fan are placed inside to convect cold air.}
    \label{figure_AVRG_temperature}
\end{figure}

In addition, we removed one of the speakers from the AVRG shown in Fig.~\ref{figure_implementation_overview} and applied electrical energy to the peltier element to create a temperature difference. Fig.~\ref{figure_AVRG_temperature} shows the AVRG. A CPU cooler and a large fan are placed outside to release heat, and a heat sink and a small fan are placed inside to convect cold air.
We believe that the AVRG can be used not to talk to DHH but to notify different information, such as the sound of running water, by emitting air vortex rings with different temperatures.

\begin{figure}[h!]
    \centering
    \includegraphics[keepaspectratio, width=1.0\textwidth]{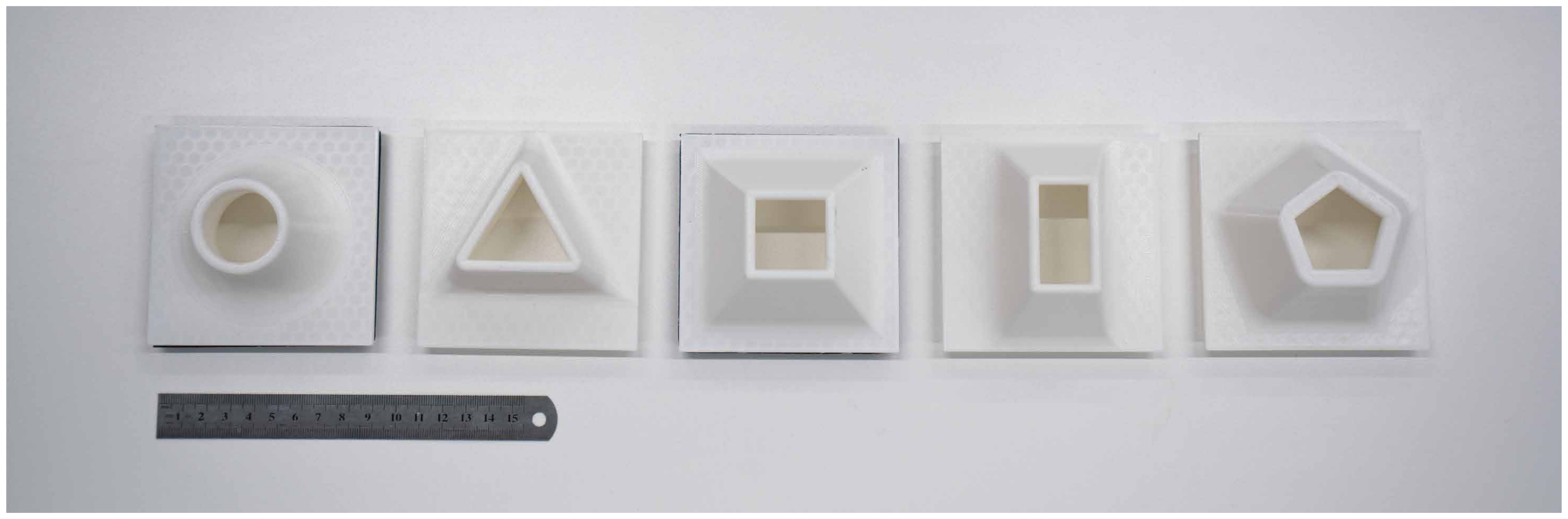}
    \caption{Triangular, square, rectangular, and pentagonal nozzles as well as circular nozzles. The shape of the air vortex ring generated by each nozzle changes during the process of translation.}
    \label{figure_nozzle_shape}
\end{figure}

As shown in Fig.~\ref{figure_nozzle_shape}, we produced not only round but also triangular, square, rectangular, and pentagonal nozzles. The shape of the air vortex ring generated by each nozzle changes during the process of translation. Therefore, not only does the shape of the vortex ring change as the nozzle shape changes but also the distance between the AVRG and the object may change the tactile sensation that can be presented.

In addition, by incorporating a module that can change the smell as well as the temperature, the auditory information can be converted into not only tactile information but also olfactory information, thereby increasing the amount of information presented.

\begin{figure}[h!]
    \centering
    \includegraphics[keepaspectratio, width=1.0\textwidth]{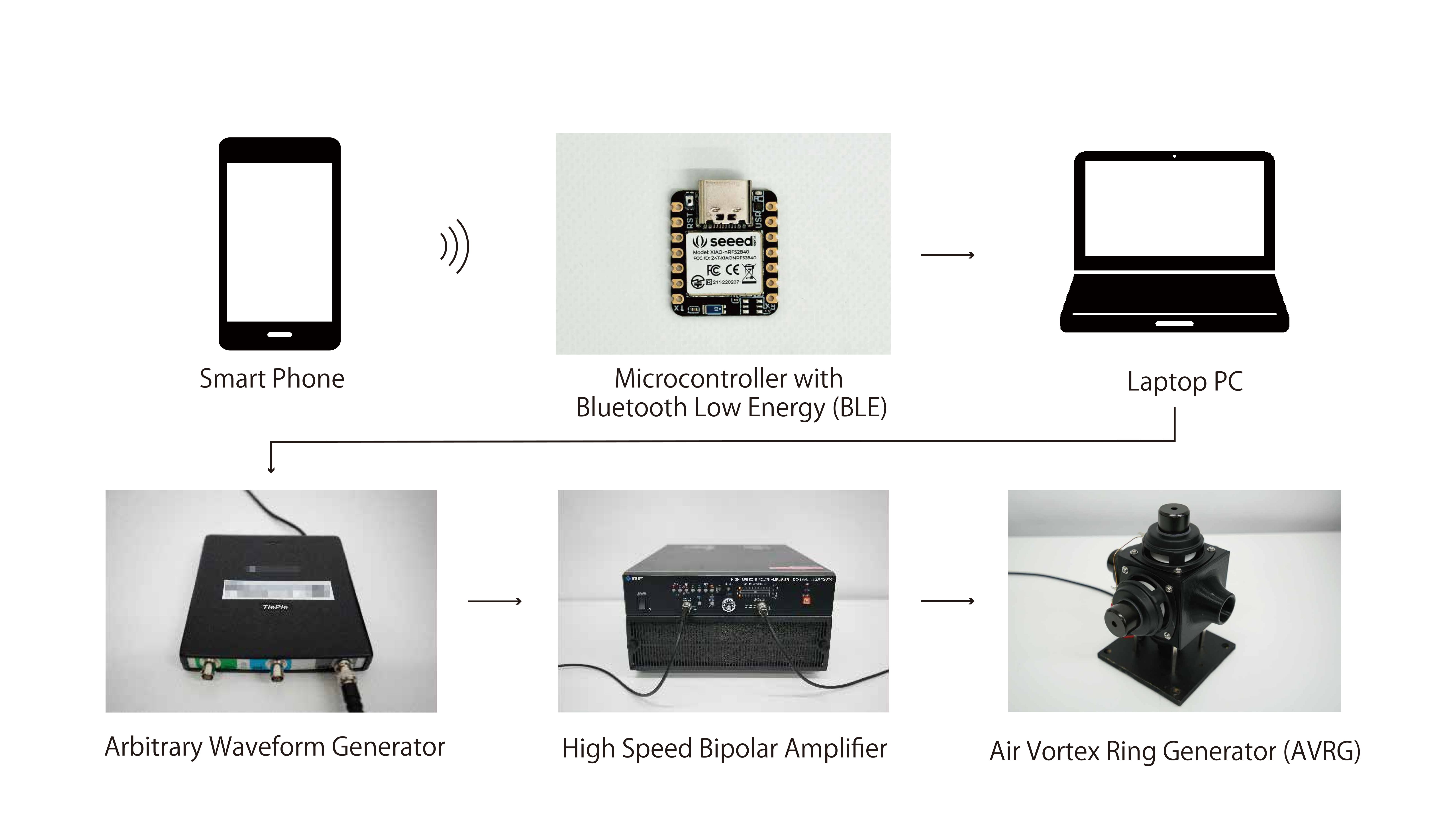}
    \caption{A system that can remotely control AVRG with a smartphone by adding a microcontroller with Bluetooth Low Energy (BLE) functionality.}
    \label{figure_system_BLE}
\end{figure}

In the system shown in Fig.~\ref{figure_implementation_overview}, the AVRG had to be operated from a laptop PC. Therefore, it was not possible to trigger the injection of AVRGs by actions such as talking to them. For this reason, we developed the system which sends voltage waveforms to the AVRG by transmitting numerical values via Bluetooth from a smartphone, as shown in Fig.~\ref{figure_system_BLE}. This system enables the AVRG to be triggered remotely by a user's operation as long as a smartphone is available, making it easier to use the AVRG for talking to a user.

\subsection{Application}
Based on the results of the accuracy experiment and user studies, we discussed how our new approach to utilizing AVRG should be designed for implementation in various situations.  

\subsubsection{Starting a conversation}
\begin{figure}[h!]
    \centering
    \includegraphics[keepaspectratio, width=1.0\textwidth]{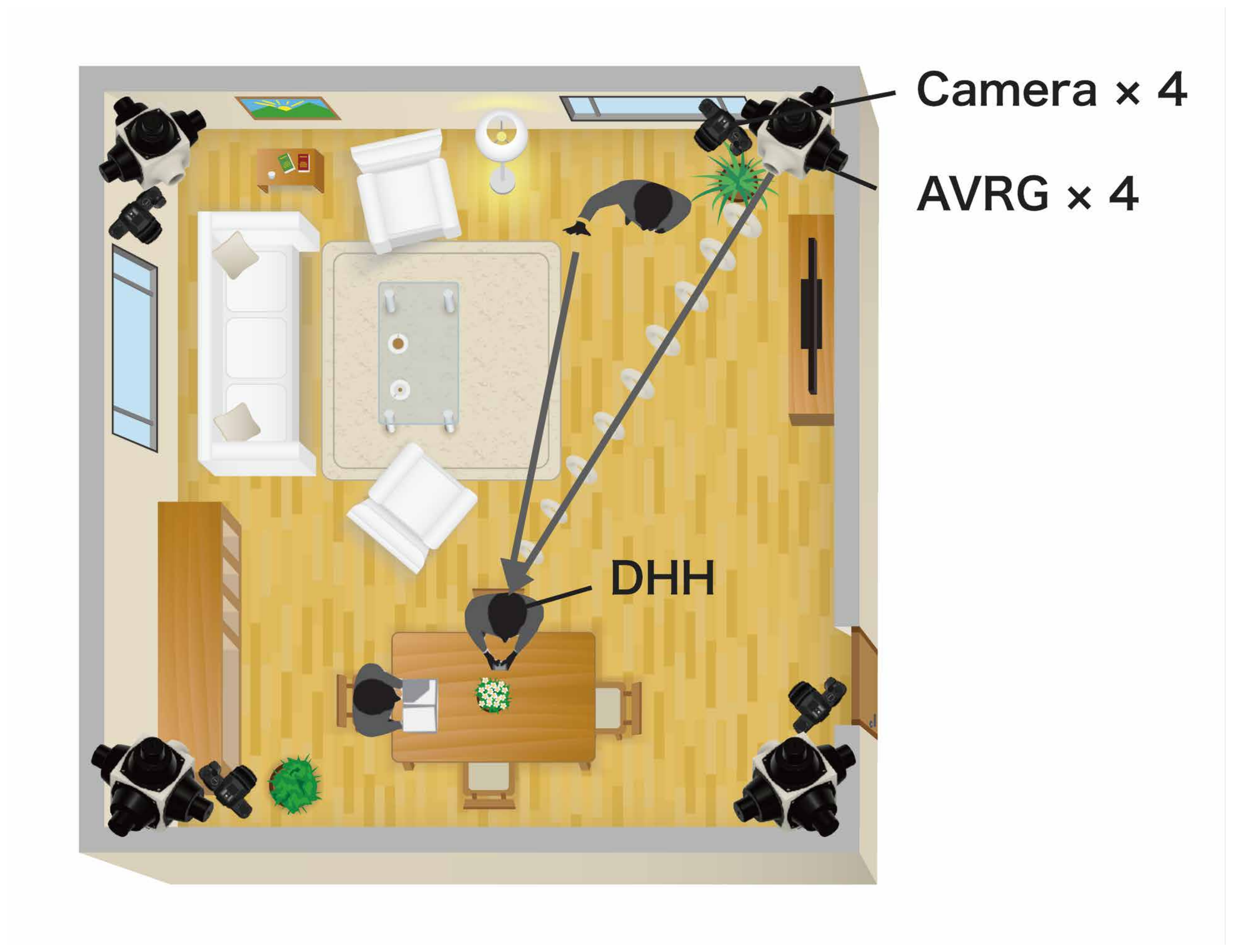}
    \caption{New approach, like lighting in a room, makes AVRGs place in a room because the DHH easier to start a conversation. Use cameras and AVRGs placed in the four corners of the room to talk to DHH using air vortex rings.}
    \label{figure_room}
  \end{figure}

 People can use any means of communication method such as being able to talk with sign language. We think it has to be mindful to enter into the DHH's field of vision, etc., to start a conversation because it is challenging for the DHH to obtain information that starts a conversation. Therefore, this new approach, like lighting in a room (Fig.~\ref{fig:teaser}(a), Fig.~\ref{figure_room}), makes AVRGs place in a room because the DHH easier to start a conversation. Specifically, image recognition using cameras and motion detection using distance sensors, etc., will detect a person waving a hand toward the opponent, and an air vortex ring is ejected with the opponent. Thus, based on user study results, we must pursue parameters related to future noticeability and comfortability of initiating conversation. In addition, we also have to consider whether waving a hand as a conversation-starting trigger is sufficient to befit the DHH.

 \subsubsection{Sound Recognition's Output and Timer, etc.}
 Considering that notifications, such as sound recognition, timers, etc., are presented in smart homes, it is necessary to devise a way to change the parameters of haptic presentation relying on the notification task instead of using a fixed haptic presentation parameter. Based on these previous studies~\cite{Goodman2020-SoundAwareness, DJ2020-TactileSoundAwareness} focusing of identifying the parameters for contact-type haptic presentation, it is necessary to investigate the method of identifying the parameters for non-contact haptic presentation using an air vortex ring.

 \subsubsection{Alert and Wake-up Alarm}
 Sounds that DHH should be aware of but is not, such as the sound of running water, a baby crying, or the sound of the intercom, abound all around us. Alerts should be considered separately from ``Sound Recognition's Output and Timer, etc.''. The reason is that the information is life-threatening and should be promptly noticed even if it is perceived as unpleasant, etc. There have been several previous studies~\cite{Judy1991-SoundDesign, Tuuri2010-SoundDesign, Guillaume2003-SoundDeSign, Christopher2021-SoundDesign} on sound, particularly in the case of hearing people. Thus, we must consider and identify the parameters for non-contact haptic presentation using an air vortex ring for alert based on reaction time and comfortability of the user study's result.
 Additionally, in the case of the wake-up alarm  ( Fig.~\ref{fig:teaser}(b)), we have to conduct experiments with DHH sleeping and consider the appropriate wake-up parameters.

 \subsubsection{Introduce into Deaf Culture and Deaf Space}
As an example of an approach that incorporates ``hand waving'' and other culturally customary acts in ``Deaf Culture''~\cite{Coates2001-Deaf, ladd2003-Deaf, padden2009-Deaf, Bartnikowska2017-Deaf}, a study of a sign language user interface~\cite{Frishberg1993-SUI, Bragg2020-SUI} is reported. In addition, the AVRGs can be used as notifications from the smart home when a DHH conversation with the smart home is based on these previous studies~\cite{Gambill2019-SUI, Glasser2020-SUI, Mande2021-SUI, Glassr2021-SUI, Kato2021-SignUI,Kato2022-SignUI}, to improve the accessibility of smart homes for the DHH.

 Moreover, we think that it is possible to apply our new approach to ''DeafSpace''~\footnote{\url{https://infoguides.rit.edu/deafspace}}\footnote{\url{https://gallaudet.edu/campus-design-facilities/campus-design-and-planning/deafspace/}}~\cite{Claire-DeafSpace}\cite{Solvang-DeafSpace} an architectural space design that considers Deaf Culture. Specifically, we are considering the possibility of doing what is described in ''Starting a conversation,'' ''Sound Recognition's Output and Timer, etc.'' and ''Alert and Wake-up Alarm'' subsections by placing them like lights everywhere in the architectural space  (Fig.~\ref{fig:teaser}(c)). Therefore, by developing a large AVRG (Fig.~\ref{figure_Large_AVRG}), we are also attempting to increase the size of the air vortex ring and strengthen which is presentation strength of the air vortex ring. However, to introduce AVRG into an architectural space, AVRG must be as small as light, where the presenting strength can be somewhat secured. This issue must be faced as a future challenge.

 \subsubsection{New Deaf Football and Futsal}
 Deaf football and futsal are modified versions of traditional football and futsal, respectively, designed for individuals with hearing impairments. These games follow the same rules as standard football and futsal, with a few adaptations to accommodate the specific needs of deaf players. In deaf football, for example, referees use flags in addition to whistles to communicate with players and provide visual cues. In international matches, there are five referees, including two assistant referees positioned behind the goals, who use flags to communicate with players. Similarly, in deaf futsal, referees use both whistles and flags to communicate with players.

 However, players do not often look at the referee when they are playing soccer and often fail to notice the referee's flags because they are looking at the ball or the opposing players. Therefore, we think that by placing huge AVRGs around the futsal field, players who fail to notice the referee's flag can be notified by air vortex rings, thereby facilitating the smooth progress of the game.
 
 In addition, when a goalkeeper wants to give instructions to a field player, there are situations in which he speaks to the player by stamping his foot and vibrating the footstep. However, it is not always possible to talk to the field players, and it may attract the attention of the opposing players. The same is true when the coach gives instructions to the players. Even if the AVRG has a short range, it has the potential to convey instructions by message if it can attract the attention of players close to the AVRG in these situations.
\section{Conclusion}
In this study, we developed a proof-of-concept device to determine the air vortex ring's accuracy and noticeability when it hits a DHH person's hair. We propose SHITARA, a novel accessibility method with air vortex rings that provides a non-contact haptic cue for a DHH person. User Study 1 shows that the air vortex ring is noticeable up to 2.5 meters away, and the optimum strength is found for each distance from DHH. In User Study 2, we applied air vortex rings to the head of DHH from seven different directions to validate how the correct rate and noticeability and comfortability of initiating conversation varied. Based on this user study's result, this method could also be used with new accessibility approaches for the DHH's daily life.
 
 In the future, we must explore presentation parameters using an air vortex ring that is a proposal notification application for DHH. Additionally, we are planning to design AVRG for on-site installation.

\section*{Financial Disclosure}
The basic research of this study was supported by Pixie Dust Technologies, Inc, and the applied research of this study was supported by JST CREST Grant Number JPMJCR19F2, Japan. We would like to thank Editage [http://www.editage.com] for editing and reviewing this manuscript for English language.

\section*{Author Contributions}
R.K., S.A., T.F., and Y.O. designed the research methodology. R.K., T.F., and A.S. contributed to the design and manufacture of AVRG system. R.K., T.F., K.T., and Y.O. designed the experiment for the noise measurement, and R.K. and K.T. collected and analyzed this experimental data. R.K., T.F., K.T., and Y.O. designed the accuracy experiment using fur, and R.K. and K.T. collected and analyzed this experimental data. R.K., S.A., T.F., R.N., K.T., and Y.O. designed the experimental environment for the user study. R.K., S.A., R.N., and K.T. collected the experimental data of the user study and R.K., S.A., and R.I. analyzed the experimental data of the user study. R.K., S.A., T.F., S.S., and Y.O. discussed the results. All authors contributed to drafting the manuscript.

\section*{Conflict of Interest}
A patent application (JP2022-131509, JP2022-171382) was filed in relation to this publication. 
\bibliographystyle{unsrt}  
\bibliography{reference} 

\begin{thebibliography}{10}

\bibitem{Suzuki_2005}
Yuriko Suzuki and Minoru Kobayashi.
\newblock Air jet driven force feedback in virtual reality.
\newblock {\em IEEE computer graphics and applications}, 25(1):44--47, 2005.

\bibitem{Iwamoto_2008}
Takayuki Iwamoto, Mari Tatezono, and Hiroyuki Shinoda.
\newblock Non-contact method for producing tactile sensation using airborne
  ultrasound.
\newblock In Manuel Ferre, editor, {\em Haptics: Perception, Devices and
  Scenarios}, pages 504--513, Berlin, Heidelberg, 2008. Springer Berlin
  Heidelberg.

\bibitem{Sodhi_2013}
Rajinder Sodhi, Ivan Poupyrev, Matthew Glisson, and Ali Israr.
\newblock Aireal: Interactive tactile experiences in free air.
\newblock {\em ACM Trans. Graph.}, 32(4), jul 2013.

\bibitem{Gupta_2013}
Sidhant Gupta, Dan Morris, Shwetak~N. Patel, and Desney Tan.
\newblock Airwave: Non-contact haptic feedback using air vortex rings.
\newblock In {\em Proceedings of the 2013 ACM International Joint Conference on
  Pervasive and Ubiquitous Computing}, UbiComp '13, page 419–428, New York,
  NY, USA, 2013. Association for Computing Machinery.

\bibitem{Sato_2017}
Yuka Sato and Ryoko Ueoka.
\newblock Investigating haptic perception of and physiological responses to air
  vortex rings on a user's cheek.
\newblock In {\em Proceedings of the 2017 CHI Conference on Human Factors in
  Computing Systems}, CHI '17, page 3083–3094, New York, NY, USA, 2017.
  Association for Computing Machinery.

\bibitem{Coates2001-Deaf}
Jennifer Coates and Rachel Sutton-Spence.
\newblock Turn-taking patterns in deaf conversation.
\newblock {\em Journal of Sociolinguistics}, 5(4):507--529, 2001.

\bibitem{ladd2003-Deaf}
Paddy Ladd.
\newblock {\em Understanding deaf culture: In search of deafhood}.
\newblock Multilingual Matters, 2003.

\bibitem{padden2009-Deaf}
Carol Padden, Tom Humphries, and Carol Padden.
\newblock {\em Inside deaf culture}.
\newblock Harvard University Press, 2009.

\bibitem{Bartnikowska2017-Deaf}
Urszula Bartnikowska.
\newblock Significance of touch and eye contact in the polish deaf community
  during conversations in polish sign language: ethnographic observations.
\newblock {\em Hrvatska Revija za Rehabilitacijska Istra $\check{z}$ ivanja},
  53:175--185, 2017.

\bibitem{World_1903}
{\em World Today}.
\newblock Number v. 5. Current Encyclopedia Company, 1903.

\bibitem{Gault_1927}
Robert~H. Gault.
\newblock “hearing” through the sense organs of touch and vibration.
\newblock {\em Journal of the Franklin Institute}, 204(3):329--358, 1927.

\bibitem{Cloud_1933}
D.~T. CLOUD.
\newblock Some results from the use of the gault-teletactor.
\newblock {\em American Annals of the Deaf}, 78(3):200--203, 1933.

\bibitem{Tan_1997}
Hong~Z Tan and Alex Pentland.
\newblock Tactual displays for wearable computing.
\newblock {\em Personal Technologies}, 1(4):225--230, 1997.

\bibitem{Tan_2005}
Hong~Z. Tan and Alex Pentland.
\newblock Tactual displays for sensory substitution and wearable computers.
\newblock In {\em ACM SIGGRAPH 2005 Courses}, SIGGRAPH '05, page 105–es, New
  York, NY, USA, 2005. Association for Computing Machinery.

\bibitem{Honda2022}
Tatsuya Honda, Tetsuaki Baba, and Makoto Okamoto.
\newblock Ontenna: Design and social implementation of auditory information
  transmission devices using tactile and visual senses.
\newblock In Klaus Miesenberger, Georgios Kouroupetroglou, Katerina Mavrou,
  Roberto Manduchi, Mario Covarrubias~Rodriguez, and Petr Pen{\'a}z, editors,
  {\em Computers Helping People with Special Needs}, pages 130--138, Cham,
  2022. Springer International Publishing.

\bibitem{DJ2020-SoundWatch}
Dhruv Jain, Hung Ngo, Pratyush Patel, Steven Goodman, Leah Findlater, and Jon
  Froehlich.
\newblock Soundwatch: Exploring smartwatch-based deep learning approaches to
  support sound awareness for deaf and hard of hearing users.
\newblock In {\em The 22nd International ACM SIGACCESS Conference on Computers
  and Accessibility}, ASSETS '20, New York, NY, USA, 2020. Association for
  Computing Machinery.

\bibitem{DJ2015-SoundAwareness}
Dhruv Jain, Leah Findlater, Jamie Gilkeson, Benjamin Holland, Ramani
  Duraiswami, Dmitry Zotkin, Christian Vogler, and Jon~E. Froehlich.
\newblock Head-mounted display visualizations to support sound awareness for
  the deaf and hard of hearing.
\newblock In {\em Proceedings of the 33rd Annual ACM Conference on Human
  Factors in Computing Systems}, CHI '15, page 241–250, New York, NY, USA,
  2015. Association for Computing Machinery.

\bibitem{Guo2020-HoloSound}
Ru~Guo, Yiru Yang, Johnson Kuang, Xue Bin, Dhruv Jain, Steven Goodman, Leah
  Findlater, and Jon Froehlich.
\newblock Holosound: Combining speech and sound identification for deaf or hard
  of hearing users on a head-mounted display.
\newblock In {\em The 22nd International ACM SIGACCESS Conference on Computers
  and Accessibility}, ASSETS '20, New York, NY, USA, 2020. Association for
  Computing Machinery.

\bibitem{Goodman2020-SoundAwareness}
Steven Goodman, Susanne Kirchner, Rose Guttman, Dhruv Jain, Jon Froehlich, and
  Leah Findlater.
\newblock Evaluating smartwatch-based sound feedback for deaf and
  hard-of-hearing users across contexts.
\newblock In {\em Proceedings of the 2020 CHI Conference on Human Factors in
  Computing Systems}, CHI '20, page 1–13, New York, NY, USA, 2020.
  Association for Computing Machinery.

\bibitem{DJ2020-HoloSound}
Dhruv Jain, Kelly Mack, Akli Amrous, Matt Wright, Steven Goodman, Leah
  Findlater, and Jon~E. Froehlich.
\newblock Homesound: An iterative field deployment of an in-home sound
  awareness system for deaf or hard of hearing users.
\newblock In {\em Proceedings of the 2020 CHI Conference on Human Factors in
  Computing Systems}, CHI '20, page 1–12, New York, NY, USA, 2020.
  Association for Computing Machinery.

\bibitem{WasabiAlarm}
Hideaki GotoTomo SakaiKoichiro MizoguchiYukinobu~TajimaMakoto Imai.
\newblock Odor generation alarm and method for informing unusual situation,
  July 2015.
\newblock Patent No. US9082274B2, Filed February 5th., 2009, Issued July. 14.,
  2015.

\bibitem{Iijima2021-SmartphoneDrum}
Ryo Iijima, Akihisa Shitara, Sayan Sarcar, and Yoichi Ochiai.
\newblock Smartphone drum: Gesture-based digital musical instruments
  application for deaf and hard of hearing people.
\newblock In {\em Symposium on Spatial User Interaction}, SUI '21, New York,
  NY, USA, 2021. Association for Computing Machinery.

\bibitem{Shitara2018-HaptStarter}
Akihisa Shitara, Miki Namatame, and Yuhki Shiraishi.
\newblock Proposal of a vibration stimulus start system for deaf and hard of
  hearing.
\newblock {\em Journal on Technology \& Persons with Disabilities}, 6:140--148,
  2018.
\newblock \url{http://hdl.handle.net/10211.3/202992}.

\bibitem{Shitara2019-HaptStarter}
Akihisa Shitara, Miki Namatame, and Yuhki Shiraishi.
\newblock Tactile stimulus start system proposal for deaf and hard of hearing
  sprinters.
\newblock In {\em Proceedings of The 7th International Conference for Universal
  Design in Bangkok 2019}, pages 185--192, Yokohama, Japan, 2019. IAUD.
\newblock The 7th International Conference for Universal Design
  \url{https://www.ud2019.net/session/en_oral-sessions.html}.

\bibitem{Goertz_1949}
Ray~C Goertz.
\newblock {\em Master-slave manipulator}, volume 2635.
\newblock Argonne National Laboratory, 1949.

\bibitem{Sheridan_1989}
T.B. Sheridan.
\newblock Telerobotics.
\newblock {\em Automatica}, 25(4):487--507, 1989.

\bibitem{Mosher_1964}
Ralph~S. Mosher.
\newblock Industrial manipulators.
\newblock {\em Scientific American}, 211(4):88--97, 1964.

\bibitem{Geldard_1966}
Frank~A Geldard.
\newblock Cutaneous coding of optical signals: The optohapt.
\newblock {\em Perception \& Psychophysics}, 1(11):377--381, 1966.

\bibitem{Bach_1969}
Paul Bach-y Rita, Carter~C Collins, Frank~A Saunders, Benjamin White, and
  Lawrence Scadden.
\newblock Vision substitution by tactile image projection.
\newblock {\em Nature}, 221(5184):963--964, 1969.

\bibitem{White_1970}
Benjamin~W White, Frank~A Saunders, Lawrence Scadden, Carter~C Collins, et~al.
\newblock Seeing with the skin.
\newblock {\em Perception \& Psychophysics}, 7(1):23--27, 1970.

\bibitem{Craig_1982}
James~C Craig and Carl~E Sherrick.
\newblock Dynamic tactile displays.
\newblock {\em Tactual perception: A sourcebook}, pages 209--233, 1982.

\bibitem{Fisher_1987}
S.~S. Fisher, M.~McGreevy, J.~Humphries, and W.~Robinett.
\newblock Virtual environment display system.
\newblock In {\em Proceedings of the 1986 Workshop on Interactive 3D Graphics},
  I3D '86, page 77–87, New York, NY, USA, 1987. Association for Computing
  Machinery.

\bibitem{Brooks_1990}
Frederick~P. Brooks, Ming Ouh-Young, James~J. Batter, and P.~Jerome~Kilpatrick.
\newblock Project gropehaptic displays for scientific visualization.
\newblock In {\em Proceedings of the 17th Annual Conference on Computer
  Graphics and Interactive Techniques}, SIGGRAPH '90, page 177–185, New York,
  NY, USA, 1990. Association for Computing Machinery.

\bibitem{Iwata_1990}
Hiroo Iwata.
\newblock Artificial reality with force-feedback: Development of desktop
  virtual space with compact master manipulator.
\newblock In {\em Proceedings of the 17th Annual Conference on Computer
  Graphics and Interactive Techniques}, SIGGRAPH '90, page 165–170, New York,
  NY, USA, 1990. Association for Computing Machinery.

\bibitem{Massie_1994}
Thomas~H Massie, J~Kenneth Salisbury, et~al.
\newblock The phantom haptic interface: A device for probing virtual objects.
\newblock In {\em Proceedings of the ASME winter annual meeting, symposium on
  haptic interfaces for virtual environment and teleoperator systems},
  volume~55, pages 295--300. Chicago, IL, 1994.

\bibitem{Iwata_2001}
Hiroo Iwata, Hiroaki Yano, Fumitaka Nakaizumi, and Ryo Kawamura.
\newblock Project feelex: Adding haptic surface to graphics.
\newblock In {\em Proceedings of the 28th Annual Conference on Computer
  Graphics and Interactive Techniques}, SIGGRAPH '01, page 469–476, New York,
  NY, USA, 2001. Association for Computing Machinery.

\bibitem{Bach_2003}
Paul~Bach y~Rita and Stephen {W. Kercel}.
\newblock Sensory substitution and the human–machine interface.
\newblock {\em Trends in Cognitive Sciences}, 7(12):541--546, 2003.

\bibitem{Ochiai_2016}
Yoichi Ochiai, Kota Kumagai, Takayuki Hoshi, Jun Rekimoto, Satoshi Hasegawa,
  and Yoshio Hayasaki.
\newblock Fairy lights in femtoseconds: Aerial and volumetric graphics rendered
  by focused femtosecond laser combined with computational holographic fields.
\newblock {\em ACM Trans. Graph.}, 35(2), feb 2016.

\bibitem{Heilig_1962}
Morton~L Heilig.
\newblock Sensorama simulator, August~28 1962.
\newblock US Patent 3,050,870.

\bibitem{Amemiya_1999}
Kenichi Amemiya and Yutaka Tanaka.
\newblock Portable tactile feedback interface using air jet.
\newblock In {\em The 9th International Conference on Artificial Reality and
  Telexistence Proceedings}, volume~99, pages 115--122, 1999.

\bibitem{Carter_2013}
Tom Carter, Sue~Ann Seah, Benjamin Long, Bruce Drinkwater, and Sriram
  Subramanian.
\newblock Ultrahaptics: Multi-point mid-air haptic feedback for touch surfaces.
\newblock In {\em Proceedings of the 26th Annual ACM Symposium on User
  Interface Software and Technology}, UIST '13, page 505–514, New York, NY,
  USA, 2013. Association for Computing Machinery.

\bibitem{Long_2014}
Benjamin Long, Sue~Ann Seah, Tom Carter, and Sriram Subramanian.
\newblock Rendering volumetric haptic shapes in mid-air using ultrasound.
\newblock {\em ACM Trans. Graph.}, 33(6), nov 2014.

\bibitem{Monnai_2014}
Yasuaki Monnai, Keisuke Hasegawa, Masahiro Fujiwara, Kazuma Yoshino, Seki
  Inoue, and Hiroyuki Shinoda.
\newblock Haptomime: Mid-air haptic interaction with a floating virtual screen.
\newblock In {\em Proceedings of the 27th Annual ACM Symposium on User
  Interface Software and Technology}, UIST '14, page 663–667, New York, NY,
  USA, 2014. Association for Computing Machinery.

\bibitem{Hashizume_2017}
Satoshi Hashizume, Amy Koike, Takayuki Hoshi, and Yoichi Ochiai.
\newblock Sonovortex: Rendering multi-resolution aerial haptics by aerodynamic
  vortex and focused ultrasound.
\newblock In {\em ACM SIGGRAPH 2017 Posters}, SIGGRAPH '17, New York, NY, USA,
  2017. Association for Computing Machinery.

\bibitem{Hoshi_2010}
Takayuki Hoshi, Masafumi Takahashi, Takayuki Iwamoto, and Hiroyuki Shinoda.
\newblock Noncontact tactile display based on radiation pressure of airborne
  ultrasound.
\newblock {\em IEEE Transactions on Haptics}, 3(3):155--165, 2010.

\bibitem{Shtarbanov_2018}
Ali Shtarbanov and V.~Michael Bove~Jr.
\newblock Free-space haptic feedback for 3d displays via air-vortex rings.
\newblock In {\em Extended Abstracts of the 2018 CHI Conference on Human
  Factors in Computing Systems}, CHI EA '18, page 1–6, New York, NY, USA,
  2018. Association for Computing Machinery.

\bibitem{Takeda_2020}
Toki Takeda, Arinobu Niijima, Takafumi Mukouchi, and Takashi Satou.
\newblock Creating illusion of wind blowing with air vortex-induced apparent
  tactile motion.
\newblock In {\em Extended Abstracts of the 2020 CHI Conference on Human
  Factors in Computing Systems}, CHI EA '20, page 1–7, New York, NY, USA,
  2020. Association for Computing Machinery.

\bibitem{Shultz_2022}
Craig Shultz and Chris Harrison.
\newblock Lrair: Non-contact haptics using synthetic jets.
\newblock In {\em 2022 IEEE Haptics Symposium (HAPTICS)}, pages 1--6, 2022.

\bibitem{Gharib_1998}
MORTEZA GHARIB, EDMOND RAMBOD, and KARIM SHARIFF.
\newblock A universal time scale for vortex ring formation.
\newblock {\em Journal of Fluid Mechanics}, 360:121–140, 1998.

\bibitem{Krueger_2005}
PAUL~S. KRUEGER.
\newblock An over-pressure correction to the slug model for vortex ring
  circulation.
\newblock {\em Journal of Fluid Mechanics}, 545:427–443, 2005.

\bibitem{Krueger_2008}
Paul~S. Krueger.
\newblock Circulation and trajectories of vortex rings formed from tube and
  orifice openings.
\newblock {\em Physica D: Nonlinear Phenomena}, 237(14):2218--2222, 2008.
\newblock Euler Equations: 250 Years On.

\bibitem{Yanagida_2004}
Y.~Yanagida, S.~Kawato, H.~Noma, A.~Tomono, and N.~Tesutani.
\newblock Projection based olfactory display with nose tracking.
\newblock In {\em IEEE Virtual Reality 2004}, pages 43--50, 2004.

\bibitem{Crossley_2021}
Eleanor Crossley, Tim Biggs, Phillip Brown, and Tahwinder Singh.
\newblock The accuracy of iphone applications to monitor environmental noise
  levels.
\newblock {\em The Laryngoscope}, 131(1):E59--E62, 2021.

\bibitem{Sugiura_2014}
Yuta Sugiura, Koki Toda, Takayuki Hoshi, Youichi Kamiyama, Takeo Igarashi, and
  Masahiko Inami.
\newblock Graffiti fur: Turning your carpet into a computer display.
\newblock In {\em Proceedings of the 27th Annual ACM Symposium on User
  Interface Software and Technology}, UIST '14, page 149–156, New York, NY,
  USA, 2014. Association for Computing Machinery.

\bibitem{Lee_2006}
Jin-Hee Lee, Su-Jeong Shin, and Cynthia Istook.
\newblock Analysis of human head shapes in the united states.
\newblock {\em International journal of human ecology}, 7, 01 2006.

\bibitem{DJ2020-TactileSoundAwareness}
Dhruv Jain, Brendon Chiu, Steven Goodman, Chris Schmandt, Leah Findlater, and
  Jon~E. Froehlich.
\newblock Field study of a tactile sound awareness device for deaf users.
\newblock In {\em Proceedings of the 2020 International Symposium on Wearable
  Computers}, ISWC '20, page 55–57, New York, NY, USA, 2020. Association for
  Computing Machinery.

\bibitem{Judy1991-SoundDesign}
Judy Edworthy, Sarah Loxley, and Ian Dennis.
\newblock Improving auditory warning design: Relationship between warning sound
  parameters and perceived urgency.
\newblock {\em Human Factors}, 33(2):205--231, 1991.
\newblock PMID: 1860703.

\bibitem{Tuuri2010-SoundDesign}
Kai Tuuri and Antti Pirhonen.
\newblock Communicative functions of sounds which we call alarms.
\newblock Georgia Institute of Technology, 2010.

\bibitem{Guillaume2003-SoundDeSign}
Anne Guillaume, Lionel Pellieux, V{\'e}ronique Chastres, and Carolyn Drake.
\newblock Judging the urgency of nonvocal auditory warning signals: Perceptual
  and cognitive processes.
\newblock {\em Journal of experimental psychology: Applied}, 9(3):196, 2003.

\bibitem{Christopher2021-SoundDesign}
Christopher~J. Hansen, Michael~F. Rayo, Emily~S. Patterson, Todd Yamokoski,
  Mahmoud Abdel-Rasoul, Theodore~T. Allen, Jacob~J. Socha, and Susan~D.
  Moffatt-Bruce.
\newblock Perceptually discriminating the highest priority alarms reduces
  response time: A retrospective pre-post study at four hospitals.
\newblock {\em Human Factors}, 0(0):00187208211032870, 0.
\newblock PMID: 34320859.

\bibitem{Frishberg1993-SUI}
Nancy Frishberg, Serena Corazza, Linda Day, Sherman Wilcox, and Rolf
  Schulmeister.
\newblock Sign language interfaces.
\newblock In {\em Proceedings of the INTERACT '93 and CHI '93 Conference on
  Human Factors in Computing Systems}, CHI '93, page 194–197, New York, NY,
  USA, 1993. Association for Computing Machinery.

\bibitem{Bragg2020-SUI}
Danielle Bragg, Meredith~Ringel Morris, Christian Vogler, Raja Kushalnagar,
  Matt Huenerfauth, and Hernisa Kacorri.
\newblock Sign language interfaces: Discussing the field's biggest challenges.
\newblock In {\em Extended Abstracts of the 2020 CHI Conference on Human
  Factors in Computing Systems}, CHI EA '20, page 1–5, New York, NY, USA,
  2020. Association for Computing Machinery.

\bibitem{Gambill2019-SUI}
Evan Gambill, Raja~S Kushalnagar, Jason Rodolitz, Christian Vogler, and
  Brittany Willis.
\newblock Accessibility of voice-activated agents for people who are deaf or
  hard of hearing.
\newblock {\em Journal on Technology \& Persons with Disabilities}, 7:144--156,
  2019.
\newblock \url{http://hdl.handle.net/10211.3/210397}.

\bibitem{Glasser2020-SUI}
Abraham Glasser, Vaishnavi Mande, and Matt Huenerfauth.
\newblock Accessibility for deaf and hard of hearing users: Sign language
  conversational user interfaces.
\newblock In {\em Proceedings of the 2nd Conference on Conversational User
  Interfaces}, CUI '20, New York, NY, USA, 2020. Association for Computing
  Machinery.

\bibitem{Mande2021-SUI}
Vaishnavi Mande, Abraham Glasser, Becca Dingman, and Matt Huenerfauth.
\newblock Deaf users’ preferences among wake-up approaches during
  sign-language interaction with personal assistant devices.
\newblock In {\em Extended Abstracts of the 2021 CHI Conference on Human
  Factors in Computing Systems}, CHI EA '21, New York, NY, USA, 2021.
  Association for Computing Machinery.

\bibitem{Glassr2021-SUI}
Abraham Glasser, Vaishnavi Mande, and Matt Huenerfauth.
\newblock Understanding deaf and hard-of-hearing users' interest in
  sign-language interaction with personal-assistant devices.
\newblock In {\em Proceedings of the 18th International Web for All
  Conference}, W4A '21, New York, NY, USA, 2021. Association for Computing
  Machinery.

\bibitem{Kato2021-SignUI}
Kato Takashi, Shitara Akihisa, Kato Nobuko, and Shiraishi Yuhki.
\newblock Sign laguage conversational user interfaces using luminous
  notification and eye gaze for the deaf and hard of hearing.
\newblock In {\em Proceedings of the Fourteenth International Conference on
  Advances in Computer-Human Interactions (ACHI2021)}, pages 30--36.
  International Academy, Research, and Industry Association, 2021.

\bibitem{Kato2022-SignUI}
Kato Takashi, Shitara Akihisa, Kato Nobuko, and Shiraishi Yuhki.
\newblock Notification, wake-up, and feedback of conversationala natural user
  interface for the deaf and hard of hearing.
\newblock {\em International Journal on Advances in Software}, 15(1 \&
  2):65--84, 2022.

\bibitem{Claire-DeafSpace}
Claire Edwards and Gill Harold.
\newblock Deafspace and the principles of universal design.
\newblock {\em Disability and Rehabilitation}, 36(16):1350--1359, 2014.
\newblock PMID: 24786970.

\bibitem{Solvang-DeafSpace}
Per Solvang and Hilde Haualand.
\newblock Accessibility and diversity: Deaf space in action.
\newblock {\em Scandinavian Journal of Disability Research}, 16:1--13, 01 2014.

\end{thebibliography}

\end{document}